\documentclass{aa}

\usepackage{booktabs}
\usepackage{amsmath}
\usepackage{graphicx}
\usepackage[varg]{txfonts}
\usepackage{xspace}
\usepackage{natbib}
\usepackage{microtype}
\usepackage[colorlinks=true,linkcolor=red,urlcolor=black,citecolor=blue]{hyperref}
\usepackage[switch]{lineno}
\usepackage{caption}
\usepackage{tabularx}
\usepackage{rotating}
\usepackage{footmisc}

\newcommand{\fu}{4U\,0115+63\xspace}
\newcommand{\nustar}{\textsl{NuSTAR}\xspace}

\newcommand{\rxte}{\textsl{RXTE}\xspace}

\newcommand{\swift}{\textsl{Swift}\xspace}

\newcommand{\bepposax}{\textsl{BeppoSAX}\xspace}
\newcommand{\swiftBAT}{\textsl{Swift} BAT\xspace}
\newcommand{\uhuru}{\textsl{Uhuru}\xspace}
\newcommand{\keVF}{ ``10\,keV feature''\xspace}

\defcitealias{Bissinger2020a}{B20}
\defcitealias{Mueller2013a}{M13}

\makeatletter
\renewcommand*\aa@pageof{, page \thepage{} of \pageref*{LastPage}}
\makeatother

\bibpunct{(}{)}{;}{a}{}{,}

\begin{document}
\title{Flux variability of the ``10\,keV feature'' of 4U\,0115+63}

\author{Katrin~Berger\inst{\ref{inst:remeis}} \thanks{\email{katrin.berger@fau.de}} \and
        Ekaterina~Sokolova-Lapa\inst{\ref{inst:remeis}}\and 
        Ralf~Ballhausen \inst{\ref{inst:umd},\ref{inst:gsfc}} \and     
        Aafia~Zainab \inst{\ref{inst:remeis}} \and
        Philipp~Thalhammer \inst{\ref{inst:remeis}} \and
        Nicolas~Zalot \inst{\ref{inst:remeis}}  \and
        Katja~Pottschmidt \inst{\ref{inst:gsfc},\ref{inst:CRESST}}\thanks{deceased 17 June 2025} \and
        Carlo~Ferrigno \inst{\ref{inst:UNIGE}, \ref{inst:BRERA}} \and
        Richard~E.~Rothschild \inst{\ref{inst:ucsd}} \and
        Felix~F\"urst \inst{\ref{inst:esa}} \and 
        Peter~Kretschmar \inst{\ref{inst:esa}} \and
        Joel~B.~Coley \inst{\ref{inst:howard},\ref{inst:aplab_gsfc}} \and
        Pragati~Pradhan \inst{\ref{inst:erau}} \and 
        Brent~F.~West \inst{\ref{inst:USNA}} \and
        Peter~A.~Becker \inst{\ref{inst:GMU}} \and
        Alicia~Rouco-Escorial \inst{\ref{inst:esa}} \and
        J\"orn~Wilms \inst{\ref{inst:remeis}} }
\authorrunning{Katrin\ Berger et al.}

\institute{
    Dr.\ Karl-Remeis-Observatory and Erlangen Centre for Astroparticle Physics,
    Friedrich-Alexander-Universit\"at Erlangen-N\"urnberg,
    Sternwartstr.~7, 96049 Bamberg, Germany
    \label{inst:remeis}
\and
    University of Maryland, Department of Astronomy, College Park, MD 20742, USA \label{inst:umd}
\and 
    NASA Goddard Space Flight Center, Astrophysics Science Division, Greenbelt, MD 20771, USA 
    \label{inst:gsfc}
\and
    CRESST and Center for Space Sciences and Technology, University of Maryland, Baltimore County, 1000 Hilltop Circle, Baltimore, MD 21250, USA \label{inst:CRESST}
\and
    Department of Astronomy, University of Geneva, Chemin d’Écogia, 16, 1290 Versoix, Switzerland 
    \label{inst:UNIGE}
\and
    INAF, Osservatorio Astronomico di Brera, Via E. Bianchi 46, 23807, Merate, Italy 
    \label{inst:BRERA}
\and
    Department of Astronomy and Astrophysics, University of California San Diego, La Jolla, CA 92093 USA 
    \label{inst:ucsd}
\and       
    European Space Agency (ESA), European Space Astronomy Centre (ESAC), Camino Bajo del Castillo s/n, 28692 Villanueva de la Cañada, Madrid, Spain 
    \label{inst:esa}
\and
    Howard University, Department of  Physics and Astronomy, Washington, D.C. 20059, USA 
    \label{inst:howard}
\and 
    NASA Goddard Space Flight Center, Astrophysics Science Division, Code 661, Greenbelt, MD 20771, USA  
    \label{inst:aplab_gsfc}
\and
    Embry Riddle Aeronautical University, 3700 Willow Creek Road, Prescott, AZ 86301, USA 
    \label{inst:erau} 
\and
    Physics Dept., United States Naval Academy, Annapolis, MD 21402, USA
    \label{inst:USNA}
\and
    Dept.\ of Physics and Astronomy, George Mason University, Fairfax, VA 22030, USA
    \label{inst:GMU}
}

\date{Received DATE / Accepted DATE} 

\abstract{X-ray spectra of accretion-powered X-ray pulsars can often be described using a power-law continuum with a high-energy cutoff, which might be further modified by additional spectral components.
The Be X-ray binary system 4U\,0115+63 is well known for having one of the highest numbers of detected harmonics of its cyclotron resonant scattering features (CRSFs), a pronounced spectral component known as the ``10\,keV feature,'' and quasiperiodic oscillations (QPOs) with a period of about 500\,s during outbursts.}
{The changes in count rate by a factor of two during the $\sim$500\,s QPOs allow us to probe the variation in the spectral components with flux. We study the ``10\,keV feature'' in emission, aiming to disentangle it from the broadband continuum and CRSFs and investigate its origin.}
{We focus on the flux-dependent behavior of the CRSF and its harmonics, and particularly the contribution of the ``10\,keV feature,'' as seen in the flux-resolved analysis of two \textsl{NuSTAR}\xspace observations of the 2015 outburst.}
{Comparing the flux-resolved spectra of a given observation with the respective total dataset revealed a distinct change in overall spectral shape at the position of the ``10\,keV feature'' but no comparable deviation at the energies of the harmonic CRSFs. The change associated with the ``10\,keV feature'' does not seem to involve its centroid energy, which remains constant within a given observation. We find indications for an anticorrelation between the continuum flux and the ratio of the ``10\,keV feature'' flux to the continuum flux within each observation.}
{The analysis strengthens previous claims that the ``10\,keV feature'' shows some independence from the remaining features. This result supports the interpretation that the ``10\,keV feature'' has a different formation mechanism than the continuum emission, although its origin lies within the same physical environment.}


\maketitle

\section{Introduction}
\label{sec:intro}

Be X-ray binaries (BeXRBs) are a subclass of high mass X-ray binaries (HMXBs), where a neutron star (NS) is in an eccentric orbit around a Be-type star. They exhibit luminous X-ray outbursts when matter from the Be star's decretion disk flows onto the NS. Two types of outbursts have been identified. Type~I outbursts are periodic and typically occur when the accretor approaches periastron and accretes part of the decretion disk \citep{Okazaki2013}. Type~II outbursts are more luminous, can last up to several weeks, show no dependence on orbital phase, and are probably caused by a strong misalignment between the decretion disk and the binary orbit \citep[][and references therein]{Martin2014,Martin2024}. In the course of an outburst, the mass accretion rate onto the strongly magnetized NS can vary by several orders of magnitude, making BeXRBs especially useful laboratories to study accretion physics of NSs. 

One particularly interesting system is \fu, which undergoes giant type~II outbursts that typically last one to two months every few years \citep[][and references therein, hereafter M13]{Mueller2013a}. Discovered as an X-ray source by \uhuru \citep{Giacconi1972a}, \fu consists of a NS with the Be-type companion star V635 Cas \citep{Unger1998a,Negueruela2001a} in an eccentric ($e=0.34$) orbit with a 24.3\,d orbital period \citep{Rappaport1978a}. The system is located at a distance of $5.8^{+0.9}_{-0.5}$\,kpc \citep{Bailer-Jones2021a}\footnote{Gaia DR3 quotes a geometric distance of $\sim$7.3\,kpc \citep{GaiaCollab2016a,GaiaCollab2023a}}.
The NS has a spin period of $\sim$3.6\,s \citep{Cominsky1978a}. 
\fu is well known for its uniquely high number of cyclotron resonant scattering features (CRSFs). These features are created in the strong magnetic field at the poles of a NS and represent the energy difference between the discrete Landau levels. Transitions between nonadjacent levels lead to higher harmonic lines. For \fu a total of five harmonic lines were detected in the X-ray band \citep{Santangelo1999a,Heindl2000a,Ferrigno2009a}. The fundamental line is at $\sim$12\,keV, which is low compared to most other cyclotron line sources \citep{Staubert2019a} and corresponds to a magnetic field strength of ${\sim}10^{12}$\,G -- similarly one of the lowest known among HMXB pulsars. Due to the high number of lines, modeling the continuum and lines requires very careful analysis. As discussed in detail by \citetalias{Mueller2013a} and \citet[hereafter B20]{Bissinger2020a}, when using a continuum description that yields physically meaningful parameters for the CRSF parameters, their energy is independent of the source luminosity. Based on an alternative model, other publications claim a very strong correlation between the luminosity and the CRSF energy \citep[see e.g.,][]{Nakajima2006a, Li2012a, Roy2024a}; see Sect.~\ref{Sec:SpectralAnalysis} for further details and a discussion of the continuum models.
Another prominent feature, visible in \fu's light curve, is its quasiperiodic oscillations (QPOs), with an amplitude of $\sim$2 in count rate on $\sim$500\ldots 900\,s timescales found during its 1999 outbursts  \citep{Heindl2000a} and also detected in later outbursts \citep{Roy2019a}.

Of special interest to this work is \fu's spectral complexity. Generally, the X-ray spectra of HMXBs in outburst can be well described by power-law continua with a high-energy cutoff. However, additional components at intermediate energies are often required to adequately model the spectra. A complex structure of residuals around 10\,keV is particularly common when using standard phenomenological models \citep{Coburn2002a, Manikantan2023a}. They can be effectively modeled by an additional broad Gaussian component, with a centroid energy of ${\sim}10\,\mathrm{keV}$, either in emission \citep[for example,][for Cep X-4 and Cen X-3]{Vybornov2017a, Thalhammer2021a} or in absorption \citep[see, for example,][for analysis of Vela X-1]{Fuerst2014a, Diez2022a}, commonly referred to as the ``10\,keV feature.''

\fu is a system with a particularly complex intermediate energy region due to the presence of the fundamental CRSF, along with the onset of the spectral turnover. The \keVF at around $\sim$8\,keV in emission in the spectrum of this source has been reported and discussed by, for example, \citet{Ferrigno2009a}, \citetalias{Mueller2013a}, and \citetalias{Bissinger2020a}. It remains unclear whether this feature can be considered an independent physical component or if its presence merely reflects a more complex formation of the broadband continuum that cannot entirely be accounted for with a cutoff power law. Since the luminosity dependence of the other spectral features is well established, we address this issue by studying the behavior of the \keVF at different luminosities using the count rate variations in the QPOs. Up to now no proper explanation of the origin of these QPOs is given in the literature. For example, \citet{Roy2019a} and \citet{Li2025a} show that models commonly used for explaining QPOs in LMXBs can be ruled out, such as the beat or the Keplerian frequency model, disk precession, and thermal instabilities in the disk. The modulation also cannot be explained by variations in foreground absorption. While work on the origin of the QPO is clearly needed, a possible explanation consists of variations in the mass accretion rate at the poles, even though the physical origin of the modulation is unknown.  
For this purpose we analyze two observations, each performed with the Nuclear Spectroscopic Telescope Array \citep[\nustar,][]{Harrison2013a} and the Neil Gehrels Swift Observatory \citep[\swift,][]{Troja2020a,Gehrels2004a}, taken during the 2015 outburst of \fu.

The paper is structured as follows.
In Sect.~\ref{Sec:DataAquisition}, we introduce the \nustar and \swift observations of \fu and discuss the data reduction.
Section~\ref{Sec:TimingAnalysis} presents our timing analysis, while in Sect.~\ref{Sec:SpectralAnalysis} we present the averaged spectral analysis of the \swift and \nustar data.
Section~\ref{sec:FluxresolvedSpec} covers the flux-resolved spectroscopy of the \nustar observations.
We summarize and discuss our results on the \keVF and its origin in Sect.~\ref{Sec:Discussion}.

\section{Outburst profile, data acquisition, and reduction}
\label{Sec:DataAquisition}

\begin{figure}
 \resizebox{\hsize}{!}{\includegraphics[width=\columnwidth]{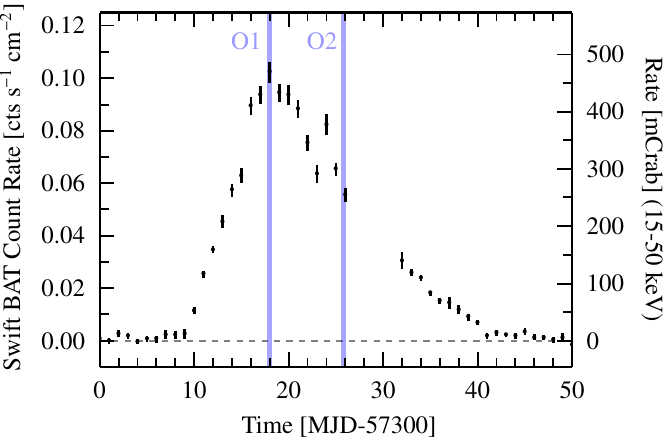}}
  \caption{\swift Burst Alert Telescope (BAT; \citet{Barthelmy2005a}) light curve of the 2015 outburst of \fu in the 15--50\,keV band with a binning of 1\,d. The blue-shaded regions highlight the quasi-simultaneous \nustar and \swift observation times.}
 \label{fig:SwiftBAT}
\end{figure}

We analyze observations taken with \nustar
 and \swift during the bright type~II outburst of \fu in 2015. This outburst started around 2015 October 15, lasted for about 30\,days, and had a peak flux of $\sim$450\,mCrab in the 15--50\,keV band of \swiftBAT (Fig.~\ref{fig:SwiftBAT}). See Table~\ref{tab:Observations} for a log of observations. Data from the 2023 outburst yield similar results to those presented here and will be discussed in a separate publication.

\begin{table}
 \centering
  \caption{List of observations of \fu.}
  \begin{tabular}{lccc}
  \hline\hline
  ID & ObsID & Start Time [UTC] & $t_\mathrm{exp}$ [s]\\
  \hline
  N1 & 90102016002 & 2015-10-22 17:30:49 & 8584 (FPMA)\\
  S1 & 00081774001 & 2015-10-23 00:17:44 & 2356 (XRT) \\
  N2 & 90102016004 & 2015-10-30 14:01:22 & 14564 (FPMA)\\
  S2 & 00081774002 & 2015-10-30 23:26:58 & 1974 (XRT) \\
  \hline
  \end{tabular}
  \label{tab:Observations}
\end{table}

\fu was observed twice by \nustar during this outburst: first on 2015 October 22 for $\sim$9\,ks (ObsID 90102016002; hereafter referred to as N1) during the outburst maximum at a 1--60\,keV luminosity of $\sim$6.8$\times10^{37}$\,erg s$^{-1}$, and on 2015 October 30 for $\sim$15\,ks (ObsID: 90102016004; hereafter called N2) at a luminosity of ${\sim}4.4\times10^{37}\,\mathrm{erg}\,\mathrm{s}^{-1}$.
Here and in the following, luminosities were determined using a distance of 5.8\,kpc \citep{Bailer-Jones2021a}.
The observation times are highlighted in Fig.~\ref{fig:SwiftBAT}. 
These two datasets have previously been analyzed by \citet{Roy2019a}, \citet{Liu2020a}, \citet{Manikantan2024a} and \citet{Stierhof2025a}, with a different focus compared to this paper.
We ignore \nustar data taken above 60\,keV, which are strongly background-dominated.
Data products were extracted using HEASOFT (version 6.30.1) and the \nustar calibration database (CALDB) (version 20220912) and reprocessed following \textsl{The NuSTAR Data Analysis Software Guide}\footnote{\url{https://heasarc.gsfc.nasa.gov/docs/nustar/analysis/nustar_swguide.pdf}}. Using newer software versions did not affect the following results.
We chose circular extraction regions of $100''$ radius centered on the source for the source region and placed background regions in the outer areas of the field of view.

To obtain a sanity check for our modeling of the soft X-ray absorption, we also utilized \swift X-ray telescope (XRT; \cite{Burrows2005a}) snapshot observations taken in window timing (WT) mode simultaneously with the \nustar observations N1 and N2, with exposure times of
$\sim$2.4\,ks (ObsID: 00081774001; hereafter S1) and $\sim$2.0\,ks
(ObsID: 00081774002; hereafter S2), respectively. Data products were extracted using HEASOFT (version 6.30.1) and \swift CALDB (version 20220803). We defined the source region with $22''\times40''$ rectangular boxes and the background region with a $25''\times45''$ area in a corner of the XRT chip.
Previous discussions of these two observations are given by \citet{Tsygankov2016a}, \citet{Wijnands2016a}, and \citet{Rouco2017a}. Since our aim is solely to check our treatment of the soft absorption, we performed a simultaneous fit to the 1.0--4.5\,keV XRT-data and the 4.5--60\,keV \nustar spectrum. These consecutive energy ranges avoid discrepancies due to known problems with cross-calibrating the two instruments \citep{Madsen2020a,Madsen2021a},  as suggested by the NuSTAR Guest observer facility (K. Pottschmidt, priv.\ comm.). Further extending the energy ranges to enable overlap between the datasets leads to comparable results but increases systematic uncertainties due to detector calibration issues. For N1/S1 the best-fit $N_\mathrm{H}$-values agree with the \nustar-only fits, while for N2/S2 the combined fit yields a slightly higher value of $N_\mathrm{H}$ at a level that does not affect our subsequent results. For this reason, and given that the XRT observations only cover a small part of the \nustar exposure time, we use \swift data only in our study of the average spectrum (Sect.~\ref{Sec:SpectralAnalysis}), not in the flux-resolved analysis (Sect.~\ref{sec:FluxresolvedSpec}).

\section{\nustar light curves and the ``500\,s QPO''}
\label{Sec:TimingAnalysis} 

\begin{figure*}
\sidecaption
  \includegraphics[width=12cm]{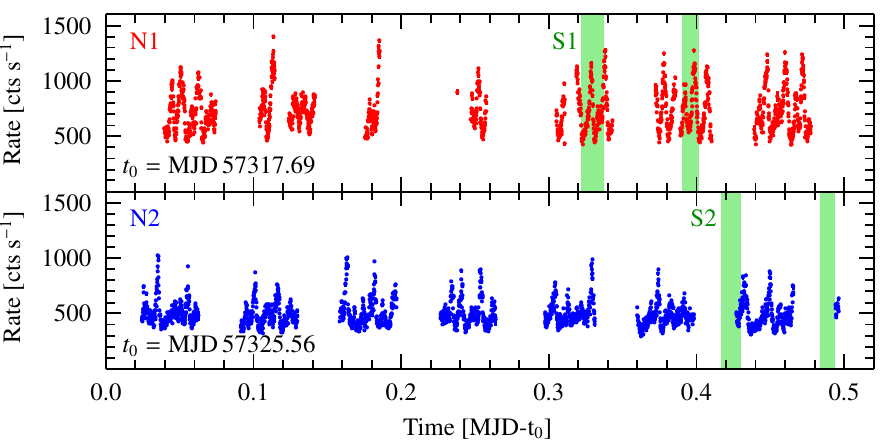}
  \caption{Light curves of N1 (top) and N2 (bottom), using the combined data from FPMA and FPMB for clarity and binned at three times the spin period of \fu. The green shaded regions highlight the observation periods of the \swift snapshots S1 and S2.}
 \label{fig:lc}
\end{figure*}

\begin{figure}
 \resizebox{\hsize}{!}{\includegraphics[width=\columnwidth]{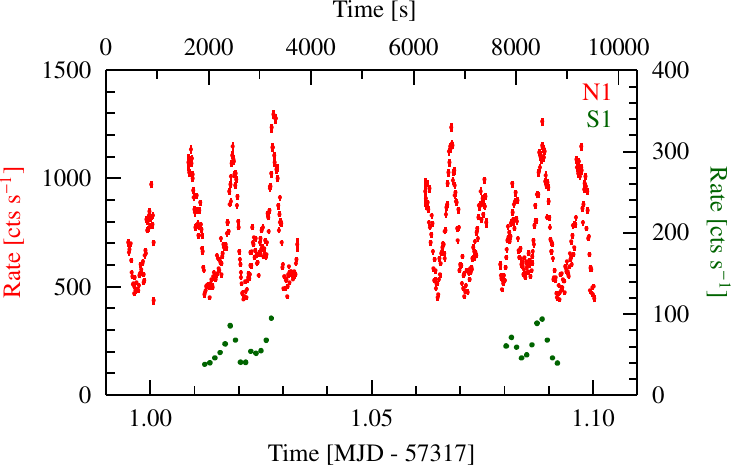}}
  \caption{Selected part of the \fu light curve, showing \nustar (red) and \swift (dark green) data highlighting the system's strong QPOs.}
 \label{fig:lc_4u1_zoom}
\end{figure}

Figure~\ref{fig:lc} shows the light curves of N1 and N2. The \swift snapshot times are highlighted by the green-shaded regions.
As previously reported by, for example, \citet{Heindl1999a}, \citet{Dugair2013a}, and \citet{Ding2021a}, the source showed a strong $\sim$2\,mHz QPO during N1 and N2, with maximum count rates of up to twice the average clearly visible on $\sim$500\,s timescales. 
Figure~\ref{fig:lc_4u1_zoom} displays a representative section of the N1 light curve and the entire S1 light curve, which spans less than four full QPO cycles.

\begin{figure}
\resizebox{\hsize}{!}{\includegraphics[width=\columnwidth]{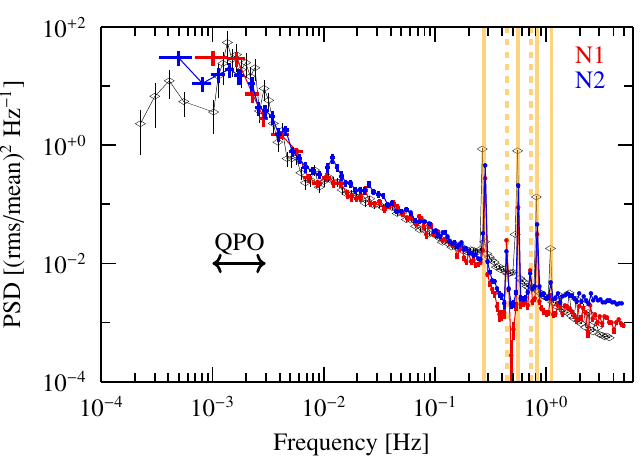}}
  \caption{PSDs of the \nustar observations, shown in red for N1 and blue for N2. For comparison the black data points indicate the results obtained by \cite{Heindl1999a}. The vertical solid orange lines highlight the spin frequency of the NS and its higher harmonics. The vertical dashed lines mark two additional peaks at 1.6 and 2.6 times the NS spin frequency.}
 \label{fig:PSD}
\end{figure}

Using power spectral densities (PSDs) from N1 and N2 observations, which
combine FPMA and FPMB, we can compare the
variability seen here with the earlier \rxte-PCA observations from
1999 \citep{Heindl1999a}. To cover a wide range of frequencies, we
utilized light curves with initial binning ranging from 0.01\,s to
100\,s across the full \nustar energy range. We used the
normalization by \cite{Miyamoto1991a}.
The PSDs are shown in Fig.~\ref{fig:PSD} superposed on the \rxte
results from \citet[][their Fig.~5, black]{Heindl1999a}. The vertical solid orange lines indicate the NS spin frequency at $\sim$0.28\,Hz and its higher harmonics. Their relative
strengths are a measure of the complexity of the pulse profile. See, for example, the PSD of Vela~X-1 discussed by \citet{Fuerst2010a}. 
Two additional smaller peaks appear at 1.6 and 2.6 times the NS's spin frequency, which are only visible in the \nustar data
(Fig.~\ref{fig:PSD}, dashed lines). The discrepancy seen above
$\sim$0.4\,Hz is due to well-known dead-time effects in the \nustar
data \citep{Bachetti2015a,Bachetti2018a}. 

Since our work focuses on low-frequencies variability, we chose not to
perform dead-time corrections to mitigate discrepancies at the high-frequency end of our dataset. At lower frequencies, the
PSDs measured with \nustar and \rxte-PCA are very similar in amplitude
and shape. In summary, \fu's variability remained remarkably
constant between the two outbursts in 1999 and 2015.
A timing analysis including the PSDs of the \nustar data from \fu's 2023 outburst was also performed by \citet{Jain2025a}. Their results are consistent with ours (except for their proper treatment of the Poisson noise correction at high frequencies, which we do not apply here). \citet{Li2025a} performed a pulse-phase resolved analysis of \fu's millihertz QPOs in the 2023 observations. They found no luminosity dependence of the frequency but report that the power of the $\sim$1\,mHz QPO is energy-dependent, while the 2\,mHz power is not.

\begin{figure*}
 \includegraphics[width=\columnwidth]{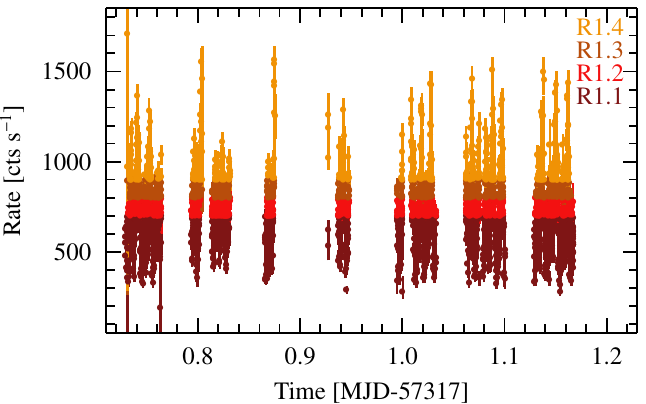}
 \includegraphics[width=\columnwidth]{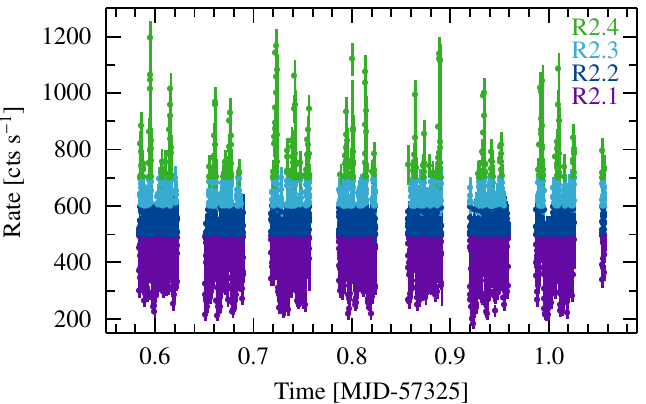}
  \caption{Selected rates for extracting different flux-resolved spectra from N1 (left) and N2 (right) using 1\,s binning. The selection criteria were chosen to be $<700$\,cts/s, 700--800\,cts/, 800--900\,cts/s, and $>900$\,cts/s for N1 and $<500$\,cts/s, 500--600\,cts/s, 600--700\,cts/s, and $>700$\,cts/s for N2.}
 \label{fig:lc_gtiRATE}
\end{figure*}

To study short term spectral changes during the QPO, we subdivided each \nustar observation into four different subsets.
To generate the good time intervals (GTIs) corresponding to a given count rate range in the full \nustar energy range, we used 1\,s binned light curves of the total observation. The selection criteria were chosen as follows. The light curve was first sliced every $n\times100\,\mathrm{counts}\,\mathrm{s}^{-1}$, where $n$ is an integer. To facilitate spectral fitting with similar signal-to-noise ratios between the spectra, we combined these slices.
The lowest count rate band required the greatest share of data with about 50\% of the total counts, as it spans the longest observation time and thus includes the highest fraction of the background counts. For both observations this resulted in more than $10^6$ counts per flux-resolved spectrum. The remaining data were then distributed approximately equally in the three higher bands. Each of the remaining count rate bands in the end held more than $2\times10^5$ counts. Initial light curve slice size affected sorting flexibility but ensured better result reproducibility and reduced the personal selection bias.
We named the flux-resolved spectra R$i.j$, where $i={1,2}$ corresponds to the N1 and N2 \nustar observations, respectively, and $j={1,2,3,4}$ denotes the serial number from the lowest to highest rate band.
The final selection of the rate bands for N1 and N2 is illustrated in Fig.~\ref{fig:lc_gtiRATE}.
During the analysis, several other intensity selections were tested, focusing, for example, on QPO phase rather than purely on count rate \citep[for details, see][]{Berger2022a}. We found that different intensity selections did not significantly change the results of the following analysis.

\section{Average spectrum and the ``10 keV feature''}
\label{Sec:SpectralAnalysis}
We now turn to the analysis of the spectra. We start with establishing
the baseline spectral shape by analyzing the flux-averaged data and use it for the flux-resolved analysis in Sect.~\ref{sec:FluxresolvedSpec}.
We modeled the joint \nustar (4.5--60\,keV) and \swift-XRT (1--4.5\,keV)
spectra using the \textsl{Interactive Spectral Interpretation System}
(ISIS), version 1.6.2-51 \citep{Houck2000a}. Both datasets were
rebinned sufficiently to warrant the applications of $\chi^2$
statistics.
Unless stated otherwise, all listed uncertainties are given at a 90\% confidence level. 

Due to the presence of the CRSFs, \fu's X-ray spectrum is notoriously difficult to model. \citetalias{Mueller2013a} and \citetalias{Bissinger2020a} give an extensive discussion of the various empirical continuum models used in the literature. They show that a high number of different combinations of continua and CRSF models (some with and without a \keVF) yield formally correct descriptions of the data in a statistical sense. The choice between modeling approaches must therefore be based on physical assumptions, not solely on fit statistics. 
\citetalias{Mueller2013a} and \citetalias{Bissinger2020a} show that continuum models based on the negative and positive exponential function
or combinations of blackbodies with a power law yield cyclotron line parameters that are not physical. For example, \citetalias{Bissinger2020a} show in their discussion of the \citet{Iyer2015a} model that the model component intended to describe the fundamental CRSF causes a relative continuum flux reduction of almost 90\% (see Fig.~5 and Table~2 from \citetalias{Bissinger2020a}). This reduction is inconsistent with our understanding of cyclotron scattering in accretion columns \citep[e.g.,][]{Schwarm2017b}. Other line parameters found with these continua, such as their very large widths, also disagree with typical CRSF parameters from other sources and expected from accretion column theory. 
For instance, \citet{Li2012a} modeled observations of the 2008 outburst with a fundamental cyclotron line as high in energy as
$\sim$17\,keV, and with higher harmonics starting at $\sim$23\,keV. Notable in this approach is its
very large width of almost 6\,keV for the fundamental line and the decreasing widths for
the higher harmonics. This behavior disagrees with
what is expected from Doppler broadening, where 
$\sigma_\mathrm{CRSF} \propto E_\mathrm{CRSF}$ \citep[][their Eq.~1]{MeszarosNagel1985b}. Applying this continuum model 
results in a CRSF energy-luminosity correlation widely cited in the literature \citep[e.g.,][]{Li2012a, Roy2024a}. However,
\citetalias{Mueller2013a} and \citetalias{Bissinger2020a} argue that this correlation appears as an artifact: the line components model changes in continuum slope and exponential cutoff with luminosity, rather than intrinsic CRSF behavior. 

In this paper, we therefore concentrate on the implications of utilizing a simple empirical continuum model that, when applied to \fu, yields CRSF parameters that are more in line with those seen in many CRSF sources \citepalias{Mueller2013a,Bissinger2020a}, namely a power law with a high energy exponential cutoff,
\begin{equation}
    N_\mathrm{ph} \propto E^{-\Gamma} \exp(-E/E_\mathrm{fold}),
\end{equation}
where $\Gamma$ is the photon index and $E_\mathrm{fold}$ the folding energy. We account for absorption in the interstellar medium using the absorption model by \citet{Wilms2000a} with cross sections by \citet{Verner1996a}.
Following \citetalias{Bissinger2020a}, we describe emission from the ionized plasma in the vicinity of the X-ray source by including three additive narrow ($\sigma = 10^{-6}$\,eV) Gaussian emission lines describing the Fe K$\alpha$ line at 6.4\,keV, the \ion{Fe}{xxv} K$\alpha$ line at 6.69\,keV, and the \ion{Fe}{xxvi} K$\alpha$ line at 6.97\,keV.
The centroids of the lines were taken from \cite{XRDB2000} and from the AtomDB Atomic Database. Above 7\,keV, the X-ray continuum is heavily modified
by cyclotron lines. We describe the lines as multiplicative absorption
lines with Gaussian optical depth profiles modeled with the
\texttt{gabs} function \citep{XspecManual}, parameterized by their
strength $D_\text{CRSF}$, width $\sigma_\mathrm{CRSF}$, and energy
$E_\text{CRSF}$. Individual lines are differentiated by consecutive
numbers, starting from 0 for the fundamental line. As we show below,
the \nustar data have a sufficiently high signal-to-noise ratio (S/N)
to detect four CRSFs.
To avoid degeneracies of the line
parameters with the continuum components, we fixed the energies of the
harmonic lines to be integer multiples of $E_\mathrm{CRSF,0}$. This
approach is justified by earlier analyses (e.g.,
\citealt{Pottschmidt2005a}, \citetalias{Mueller2013a}, and
\citetalias{Bissinger2020a}). We confirmed these results in a test
spectral fit where we set $E_{\mathrm{CRSF},n}= c_\mathrm{E} n
E_{\mathrm{CRSF},0}$ and found $c_\mathrm{E}$ to never exceed
$\pm$5\%. We therefore omitted this constant, as our analysis does not focus
on CRSF energies.

Following the argument of \citet[][their Eq.~1]{MeszarosNagel1985b}, we coupled the CRSF widths $\sigma_{\mathrm{CRSF},n}$ to the respective line energy, using the constant $c_\mathrm{CRSF, \sigma}$. This approach mitigates nonphysical correlations between the CRSFs and the continuum introduced when the parameters are left free \citepalias{Mueller2013a,Bissinger2020a}.
Finally, we also included a multiplicative constant to account for uncertainties in the relative flux calibration of the instruments, normalizing all fluxes to that measured with \nustar-FPMA. 

Other analyses, for example \citetalias{Mueller2013a}, used the \texttt{cyclabs} function instead of \texttt{gabs} to model the CRSFs. A detailed explanation of both functions is given in \cite{XspecManual}. Fits to \fu with both CRSF models were compared and discussed by \citetalias{Bissinger2020a}, who
showed that both models can describe the spectra of \fu equally well.
Therefore, the only benefit of choosing a specific model is comparability with previous results using the same model.

Consistent with earlier work, our analysis shows that a broad \keVF is
required \citep[for example,][\citetalias{Mueller2013a,Bissinger2020a}]{Ferrigno2009a,Liu2020a}.
Following \citetalias{Mueller2013a}, we added a Gaussian line profile to account for the presence of the ``10\,keV feature,'' with width $\sigma_\mathrm{10keV}$, energy
$E_\mathrm{10keV}$, and flux $F_\mathrm{10keV}$.
In another approach \citet{Farinelli2016a} modeled the \keVF using cyclotron emission convolved with a varying magnetic field in the accretion column.
\citetalias{Bissinger2020a} discuss the possibility of using a
blackbody component to model the \keVF in \fu instead of a broad Gaussian, concluding 
that it does not provide a sufficient
description of the component and leads to strong residuals below
10\,keV. We can confirm this conclusion with our own unsuccessful
attempts.

\begin{figure*}
 \includegraphics[width=\columnwidth]{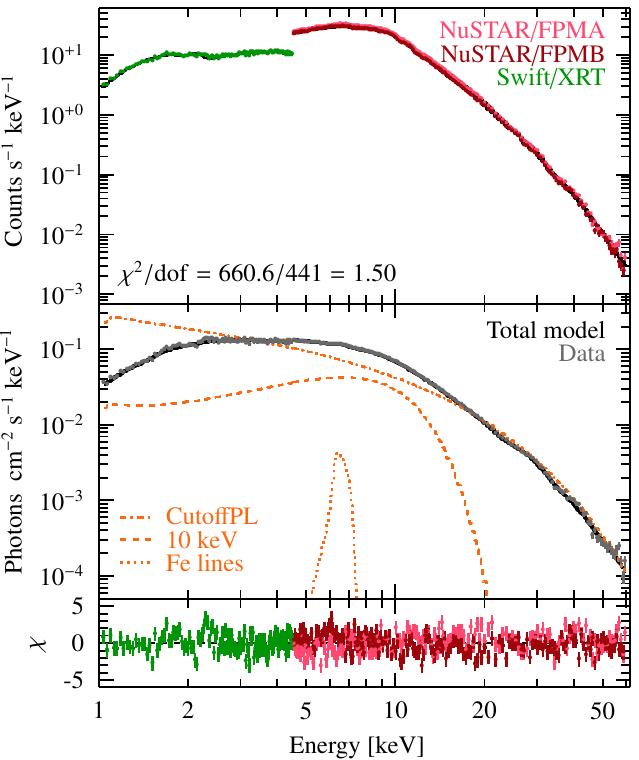}
 \includegraphics[width=\columnwidth]{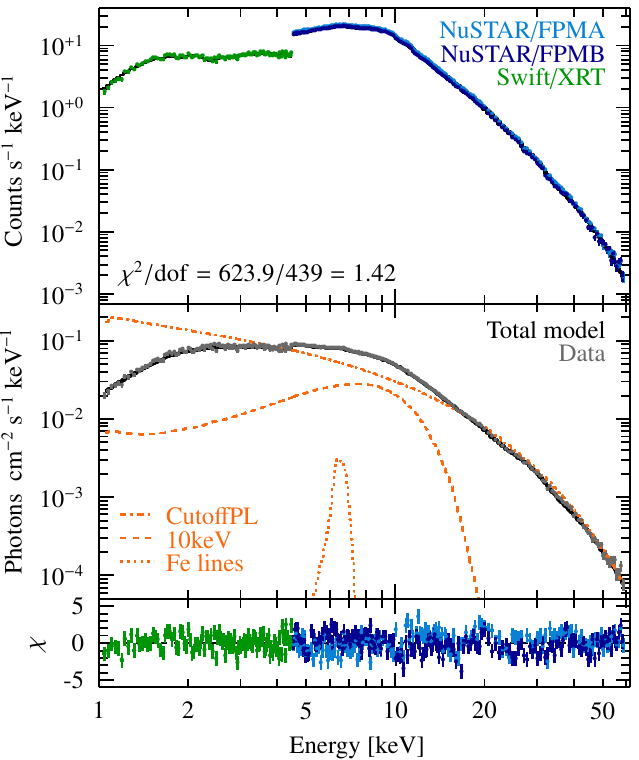}
  \caption{Top: \nustar and \swift spectra of O1 (left) and O2 (right)
    with the best-fit model (black). The data from XRT are shown in
    green, FPMA in the lighter color, and FPMB in the darker color. Middle:
    Visualization of different model components. The data are shown in
    gray, and the model using the best-fit parameters is shown in
    black. The component contribution is shown in orange,
    using a dash-dotted line for the cutoff power law, a dashed line
    for the ``10\,keV feature,'' and a dotted line for the iron lines. Bottom:
    Residuals of the best-fit model.}
 \label{fig:spec_NuSw_Comp}
\end{figure*}

Finally, we modeled the spectra using a power law with a
high energy exponential cutoff, four CRSFs, a broad Gaussian to account for the ``10\,keV feature,'' and three narrow Gaussian features for the iron lines. The resulting best-fit parameters for the two sets of combined data
from \nustar and \swift are listed in
Table~\ref{tab:totalPar_NuSTAR_Swift}; see
Fig.~\ref{fig:spec_NuSw_Comp} for the spectra and the respective best
fit. The model describes the underlying spectra well, showing no
systematic deviations in the residuals, resulting in a fit with
$\chi^2 / \mathrm{dof} = 660.6 / 441 =1.50$ for O1 and $\chi^2 /\mathrm{dof} = 623.9 / 439 =1.42$ for O2. 
The reduced $\chi^2$ can in principle be further improved when relaxing the constraints on the CRSF parameters, for example by not coupling the line energies. However, as the additional constraints were required for the subsequent flux-dependent analysis, we maintained them to ensure comparability. 

\section{Flux-dependence of the ``10\,keV feature''}
\label{sec:FluxresolvedSpec}
In this section, we use \fu's QPOs to study the behavior of its spectral components, primarily the ``10\,keV feature,'' at different flux levels on short time scales. To this end, we divided each observation based on the \nustar count rate (see Sect.~\ref{Sec:TimingAnalysis}) and conducted a spectral analysis on the flux-selected spectra. We maintained the assumptions from Sect.~\ref{Sec:SpectralAnalysis} for modeling the observation-averaged spectra, to ensure consistent fit approaches and perform the analysis solely based on the \nustar data due to lack of \swift coverage.

\subsection{Flux-resolved spectroscopy}
\label{Sec:Rates}

\begin{figure*} 
 \includegraphics[height=21\baselineskip]{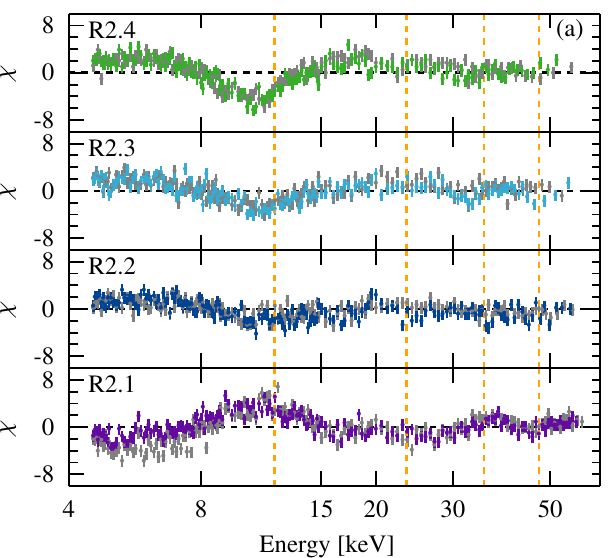}\hfill
\includegraphics[height=21\baselineskip]{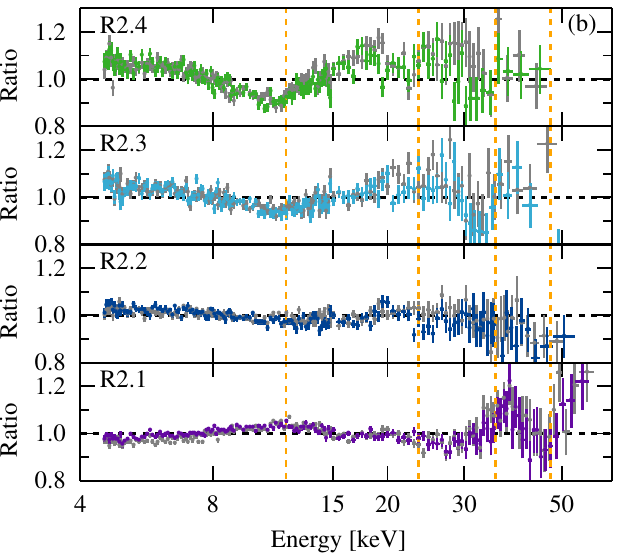}\\
 \includegraphics[height=21\baselineskip]{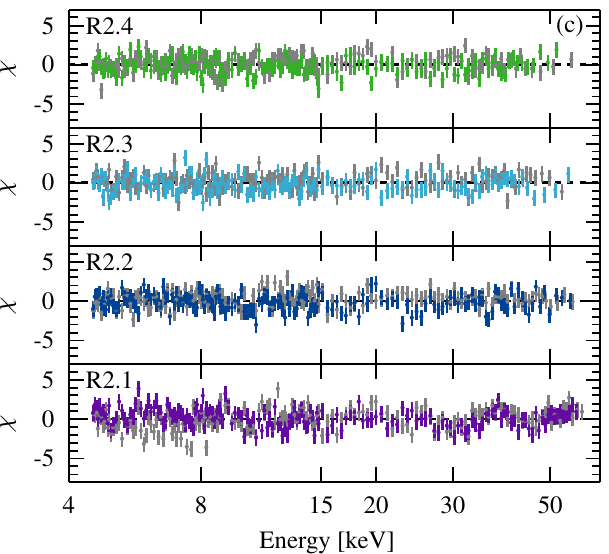}\hfill
\includegraphics[height=21\baselineskip]{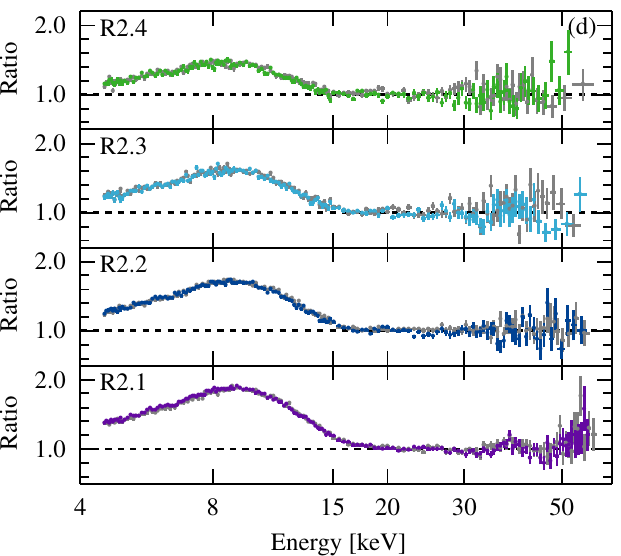}
  \caption{Residuals of the N2 flux-resolved spectra (FPMA in gray; FPMB in varying colors). The dashed orange lines mark the CRSF positions, using the values listed in Table~\ref{tab:totalPar_NuSTAR_Swift}. 
  Renormalizing the spectral model to the source flux without changing spectral parameters results in the $\chi^2$ residuals (a) and ratio (b). Refitting the parameters of the four flux-resolved spectra simultaneously leads to the best fit (c). (d): Contribution of the ``10\,keV feature,'' showing the resulting ratio after removing the component from the best-fit.}
 \label{fig:4u1_residuals}
\end{figure*}

We extracted flux-resolved \nustar spectra following the procedures outlined in Sect.~\ref{Sec:SpectralAnalysis}. Figures~\ref{fig:4u2_residuals}a and \ref{fig:4u1_residuals}a show the residuals between these flux-resolved spectra and the shape of the best-fit model from the respective time-averaged observation after a simple renormalization of the model. 
The residuals in Fig.~\ref{fig:4u2_residuals} and \ref{fig:4u1_residuals} clearly show a strong deviation from the best-fit spectral shape of the time-averaged spectrum for both observations, only around 10\,keV. For the lowest count-rate band, we clearly see a flux excess, which then transitions to a flux deficit in the highest band. Simultaneously, the spectral shape outside this band, including the harmonic CRSFs, does not vary in shape.
This result implies that, when using our continuum modeling approach, a separate spectral component may produce the flux in this band, varying less during the QPO than the remaining continuum. 

\begin{figure} 
 \includegraphics[width=\columnwidth]{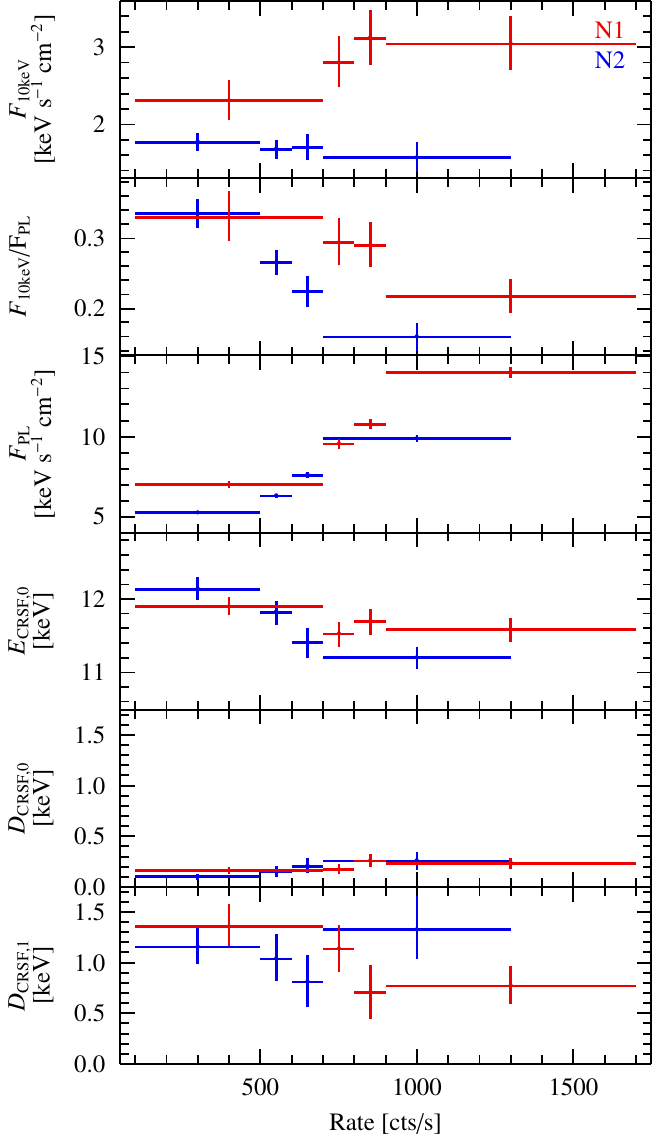}\hfill
  \caption{Selected best-fit parameters of the flux-resolved
    spectra of N1 (red) and N2 (blue), as given in
    Table~\ref{tab:RATE_Par_4ux_compact_20240516}. The horizontal bars indicate the band for the rate selection (using the full \nustar energy range). The vertical error bars show the 90\% confidence interval for the values shown.}
 \label{fig:ParHistComb_20240516}
\end{figure}

To quantify the change in the 10\,keV band, we modeled the
flux-resolved spectra with our baseline model. Since foreground absorption
is flux-independent on the short timescales considered here and \swift coverage is insufficient (covering fewer than four QPO cycles; see Fig.~\ref{fig:lc_4u1_zoom}) in our spectral fits, we couple $N_\mathrm{H}$ across the flux-resolved spectra. Similarly, several other parameters are flux-independent, including the detector constant $c_\mathrm{FPMB}$ and the coupling constant between the cyclotron line width and energy, $c_\mathrm{CRSF,\sigma}$. As a sanity check, we also performed the following analysis without coupling these constants, with similar principal results.
The energy of the \keVF was also coupled, as initial spectral fits with free $E_\mathrm{10keV}$ showed no significant variations. Applying these constraints helped us constrain the remaining spectral parameters, in particular the position of the fundamental CRSF of R1.4, as this dataset has the
lowest signal-to-noise ratio among all subsets. The resulting
simultaneous fit of all four spectra describes the spectra and
their variation well ($\chi^2 / \mathrm{dof} = 1429.2 / 1220 = 1.17$
for N1 and $\chi^2 / \mathrm{dof} = 1416.4 / 1211 = 1.17$ for N2). The
resulting best-fit parameters are listed in
Table~\ref{tab:RATE_Par_4ux_compact_20240516} and visualized  in
Fig.~\ref{fig:ParHistComb_20240516}. As shown by the residuals in
Fig.~\ref{fig:4u2_residuals}c and~\ref{fig:4u1_residuals}c, the model gives a good description of the
data. Fitting the datasets individually, instead of simultaneously as
mentioned above, leads to comparable parameters within the uncertainties.

We also attempted spectral fitting with fewer constraints, for example, individually determining $c_\mathrm{CRSF,\sigma}$ and $E_\mathrm{10keV}$ for each spectral subset. This resulted in fit statistics nearly as good as our final results; however, it produced a physically unreasonable combination of parameters, such as the fundamental CRSF describing an unreasonably high continuum fraction. In particular, the data from the lowest rate band could not be well constrained. Since our previous modeling approach already describes the data well, a more complicated model is not justified. We thus adopted it as our best-fit model.

\subsection{The ``10\,keV feature'' as a distinct spectral component}
\label{Sec:ConfidenceContours}

To visualize the overall contribution of the \keVF to the spectra, in
Figs.~\ref{fig:4u2_residuals}d and \ref{fig:4u1_residuals}d, we show
the residuals when setting the flux of the Gaussian component to zero.
The residuals clearly show that the feature has a constant width, but
that its relative contribution to the remaining continuum decreases
with increasing count rate. We note that the energy range affected by the \keVF also contains the fluorescent iron lines. Since
the lines are much narrower than the ``10\,keV feature,'' it is possible to
separate the behavior of the fluorescent lines from that of the ``10\,keV feature.'' 

Although we can separate the fluorescent lines from the feature, the
energy range around $\sim$10\,keV contains contributions from multiple
other spectral components, such as the fundamental CRSF. Degeneracies
between the parameters describing these features can complicate the
interpretation of our fit results. We therefore examine the
correlations between these components by looking at the confidence
contours for different pairs of parameters, specifically the
``10\,keV feature,'' fundamental CRSF, and cutoff power-law continuum \citep{Lampton1976a}.

\begin{figure*} 
 \includegraphics[height=42\baselineskip]{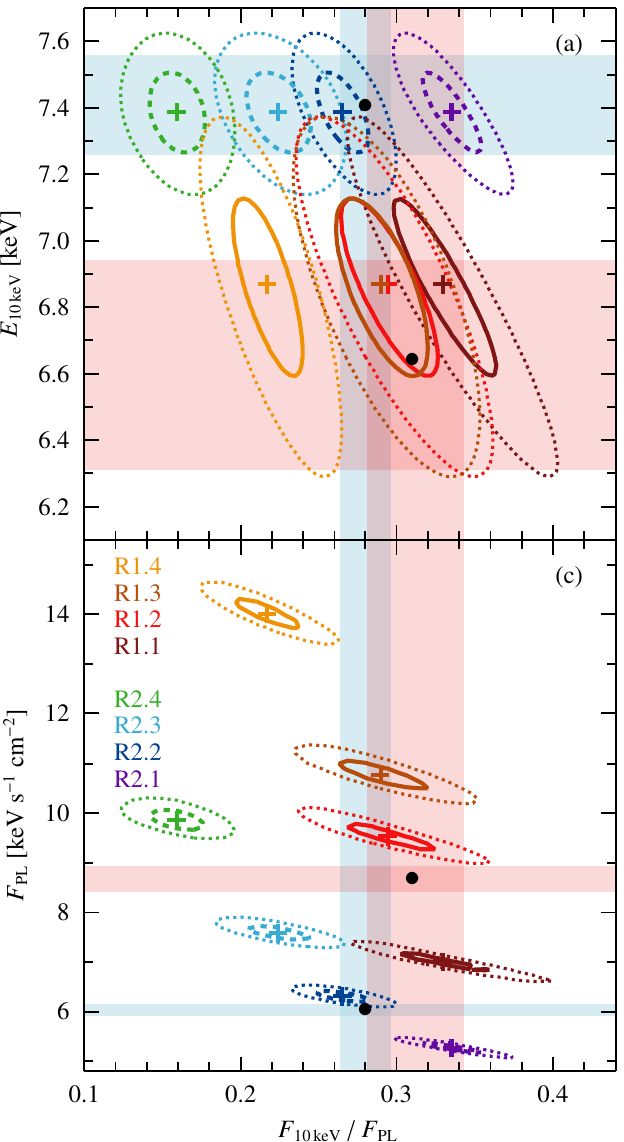}
 \includegraphics[height=42\baselineskip]{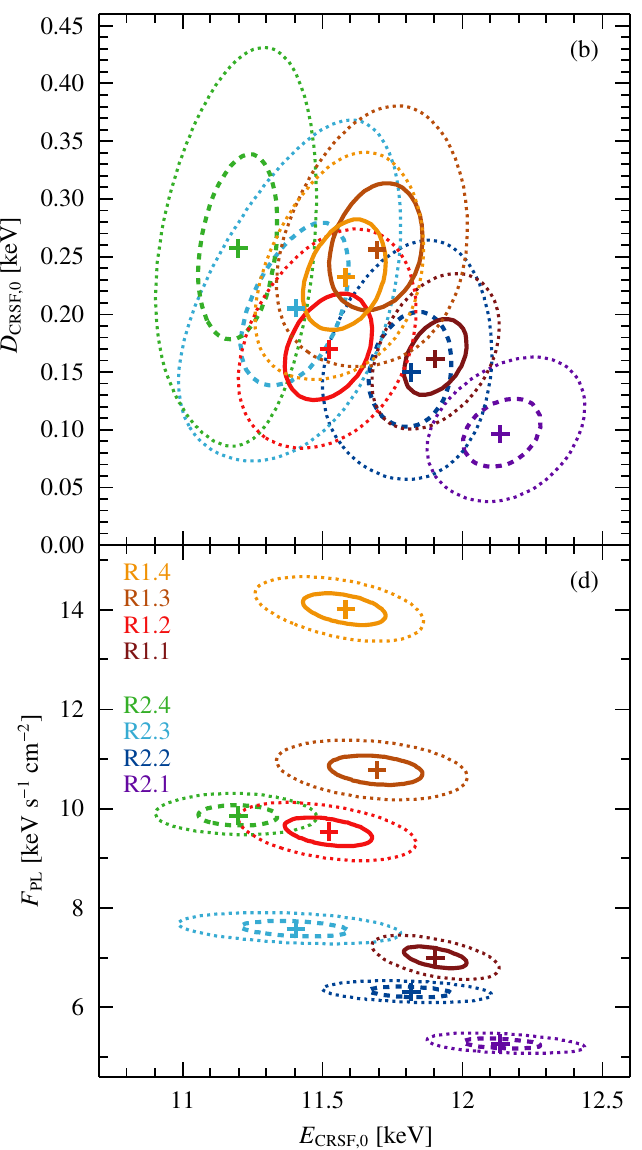} 
  \caption{Confidence contours of ratios
    between fundamental CRSF parameters, \keVF, and continuum
    flux correlations.
    For all flux-resolved spectra, the best-fit
    values are marked with a cross. The dotted lines show the $3\sigma$
    contours, and the solid lines indicate the $1\sigma$ contours for
    the N1 (red) and N2 (blue) observations,
    respectively. The black points mark the best-fit values from the
    flux-averaged dataset, while the shaded regions in the left
    panels (N1 in red and N2 in blue) highlight the uncertainties of
    the averaged best fit, as given in Table~\ref{tab:totalPar_NuSTAR_Swift}.}
 \label{fig:cm_R}
\end{figure*}

Figure~\ref{fig:cm_R} shows the confidence regions between the
relative flux of the ``10\,keV feature,'' i.e., the ratio of the flux to the
\keVF and the 3-50\,keV cutoff power-law flux,
$F_\mathrm{10keV}/F_\mathrm{PL}$, and the centroid energy of the
\keVF $E_\mathrm{10keV}$. The relative flux of the \keVF with
respect to the continuum flux increases as
continuum flux decreases, leading to separated
solutions.
Using the approach outlined by \citet{Ferrigno2020a}, which computes the distribution of correlation coefficients between two variables assuming the posterior intervals have Gaussian distributions, we find that the false alarm probability for the correlation between $F_\mathrm{10keV}/F_\mathrm{PL}$ and $F_\mathrm{PL}$ is $0.11$ for N1 and $0.05$ for N2. The corresponding $r^\mathrm{2}$ values are [0.82, 0.89, 0.94] (N1) and [0.93, 0.95, 0.97] (N2).

We emphasize that our fits coupled
$E_\mathrm{10keV}$ across the rate bands, since initial fits
showed no indication count-rate dependence on the feature's center energy.
We also emphasize that $E_\mathrm{10keV}$ changed between the two observations.

Again using confidence contours, we checked for possible correlations
between the CRSF and continuum parameters
(Fig.~\ref{fig:cm_R}b and~d). The data are described using the
same value for $c_\mathrm{CRSF,0}$ for all resolved spectra of an
observation (Table~\ref{tab:RATE_Par_4ux_compact_20240516}). For
observation N1, Fig.~\ref{fig:cm_R}b shows no visible changes in the
CRSF parameters $D_\mathrm{CRSF,0}$ and $E_\mathrm{CRSF,0}$. We note a
minor anticorrelation in the N2 analysis, but all values
for $D_\mathrm{CRSF,0}$ still agree within $3\sigma$ and provide no
evidence of flux-dependent variability, while $E_\mathrm{CRSF,0}$
only varies slightly.  
Figures~\ref{fig:cm_R}c and~d clearly show $F_\mathrm{PL}$ increasing with count rate across the selected
bands of both observations, enabling us to probe the behavior of the \keVF at different luminosities.

\section{Discussion}
\label{Sec:Discussion}
In the following we discuss the interplay between the source's CRSFs, its \keVF (Sect.~\ref{Sec:Disentangling}), and their parameter changes between datasets (Sect.~\ref{Sec:FluxCorrelations}). We conclude this section by
giving an overview of possible origins of the \keVF (Sect.~\ref{Sec:PossibleOrigin}).

\subsection{Disentangling the \keVF and the CRSFs}
\label{Sec:Disentangling}
As shown in Sect.~\ref{sec:FluxresolvedSpec}, the flux-averaged model still provides a rather good description of
the spectral shape below $\sim$6\,keV and above $\sim$20\,keV after renormalization (Fig.~\ref{fig:4u1_residuals}a). A separate spectral component, with a different flux-dependent behavior than that of the underlying continuum, must contribute to the emission at around 10\,keV.

Apart from the power-law continuum, the spectrum around 10\,keV is
shaped by two spectral components, the fundamental CRSF and the ``10\,keV feature.''
We can assess fundamental cyclotron line variability based
on the behavior of the higher harmonics, which are easier to decouple from the continuum. To investigate these changes, we look at the data-model ratios of the spectral subsets, using the rescaled
flux-averaged model. Apart from the strong deviation around
$\sim$10\,keV, we find further variations at higher energies that are less statistically significant (see Figs.~\ref{fig:4u2_residuals}a
and~\ref{fig:4u1_residuals}a). Although these deviations seem to
appear where the higher harmonic CRSFs are located, they do not
exactly align with the harmonic energies. Furthermore, the
residuals around 10\,keV can be resolved after refitting, while higher-energy wiggles remain visible (see Figures~\ref{fig:4u2_residuals}c
and~\ref{fig:4u1_residuals}c). 
This speaks against a systematic shift
in the cyclotron energy or in the strength of the higher harmonics. Moreover,
variability of the cyclotron energy with count rate
(Fig.~\ref{fig:cm_R}d) is only visible for the N2 observation, not in the
N1 data, and is therefore unlikely to be the cause of the
flux-dependence seen in both observations.

We conclude that even if the cyclotron line is slightly variable, this variability is not sufficiently strong to describe the observed flux
changes in the 10\,keV band. This is further supported by the failure
of attempts to model the rate-dependent flux change in the
$\sim$10\,keV region solely by varying the CRSF parameters and keeping both the \keVF and the continuum fixed. This leaves only the \keVF as the source of the observed flux-dependence.

\subsection{Flux-correlations of the \keVF within and between observations}
\label{Sec:FluxCorrelations}
In the following we examine the changes in the \keVF
parameters. First, the confidence contours shown in
Fig.~\ref{fig:cm_R}c again support our main result that within each
observation the flux ratio $F_\mathrm{10keV}/F_\mathrm{PL}$ is
strongly anticorrelated with the overall continuum $F_\mathrm{PL}$.
This is the first time that this flux-dependent variation in the \fu spectra is reported. In previous analyses, when comparing
multiple observations from the same or several outbursts, the change
in the spectra was not seen. It might be that this independent
behavior of the \keVF only occurs on shorter timescales
comparable with luminosity changes in the $\sim$500\,s QPOs. 
Overall, these results again hint
toward the \keVF being an additional spectral component.
This correlation is strongest in the lower-luminosity observation N2.
There is a similar trend between $F_\mathrm{PL}$ and the cyclotron
line energy $E_\mathrm{CRSF,0}$, with a more
apparent anticorrelation for the N2 observation (Fig.~\ref{fig:cm_R}d). The similarity between the two
observed correlations may point to a common origin.

However, while we do not observe any dependence of the \keVF energy on
the individual rate bands within each observation, we do find that the
centroid energy differs between the two observations, with the more
luminous observation N1 showing a lower energy for the ``10\,keV feature.'' This
trend between observations is consistent with
\citetalias{Mueller2013a}, who suggested that the centroid energy
decreases with increasing overall flux. Similarly, we
find a small variation in the feature's width and no change in
the CRSF energy between the two observations. Such a shift in the
\keVF between observations could hint at an additional luminosity
dependence of its underlying creation mechanism. This
diverging behavior for the \keVF energy and cyclotron line
rather suggests unrelated underlying physical mechanisms, in
contrast to the similar rate dependence seen within each observation.

\subsection{Possible physical origin of the ``10 keV feature''}
\label{Sec:PossibleOrigin}
Our results, particularly the anticorrelation between the flux ratio $F_\mathrm{10keV}/F_\mathrm{PL}$ and the power-law flux presented in Sect.~\ref{sec:FluxresolvedSpec}, support the interpretation that the \keVF represents a spectral component that is to some degree different in its formation from the overall continuum. However, this decoupling is not complete, as the flux of the \keVF still increases with the power-law flux, suggesting that the \keVF likely originates from the same physical environment. Based on our current understanding of spectral formation in highly magnetized plasma near the NS surface, several possible explanations for the origin of this feature can be outlined. Here, we review the proposed physical mechanisms for the formation of the spectrum at intermediate energies -- and, more specifically, the \keVF -- and present our interpretations with respect to the observed behavior.

The X-ray continuum in the accretion channel is expected to form by thermal and bulk Comptonization of bremsstrahlung, blackbody radiation, and cyclotron emission \citep[see, for example,][]{Meszaros1992, Arons1987a}. Cyclotron emission occurs when collisions between electrons and protons in a magnetized plasma excite electrons to higher Landau levels. Excitations are followed by radiative decay, producing photons with energies approximately equal to the local cyclotron energy. When collisions occur due to the thermal motion of particles, this process is known as cyclotron cooling. When the cyclotron energy is comparable to the plasma temperature, it can become the dominant cooling mechanism. Depending on reprocessing of cyclotron photons by the surrounding plasma, they can significantly contribute to the total flux. It is important to note, however, that generally all radiative processes, including bremsstrahlung and Compton scattering are affected by the strong magnetic field, which introduces anisotropy, resonances, and polarization effects.

The physical models currently available for fitting the spectra of X-ray pulsars are limited to continuum models that describe polarization- and angle-averaged bulk and thermal Comptonization of seed bremsstrahlung, blackbody, and cyclotron photons \citep{Becker2007a, Farinelli2016a}. In these models, bremsstrahlung is assumed to follow the classical, nonmagnetic cross section, while cyclotron emission accounts for magnetic effects. Such continuum models typically provide a smooth power-law-like spectrum with a high-energy cutoff dominated by the Comptonized bremsstrahlung \citep[see, e.g.,][]{Becker2007a}.
\citet{Ferrigno2009a} adopted the \citet{Becker2007a} model to fit \bepposax data of the 1999 outburst of \fu. While their phenomenological model required significant contribution from the \keVF to the total flux, the physical model was able to describe the spectra with only a minor Gaussian component at ${\sim}9\,\mathrm{keV}$ to
flatten the residuals. At the same time, the continuum formation in
the model was dominated by cyclotron emission. This was possible,
however, only by assuming a magnetic field for the cyclotron emission
almost twice as low as that estimated from the observed fundamental
cyclotron resonance. \citet{Ferrigno2009a} interpreted this discrepancy as possible evidence for a spatial separation between the region where the cyclotron emission is produced and the region where the cyclotron lines are imprinted onto the spectrum. In this case, the cyclotron emission would originate higher up in the accretion channel and then be advected downward by the bulk flow, reprocessed by the thermal plasma, and emitted from the column walls where CRSFs form. See \citet{Becker2022a} and \citet{Li2024a} for further extensions of this model and \citet{West2017a,West2017b,West2024a} for a discussion of a variant of this model where the height-dependent structure is solved using a numerical approach. 

We note, however, that \citet{Becker2007a}
assume the seed photons are mainly thermal cyclotron emission (in addition to those from bremsstrahlung and blackbody radiation), which is unlikely to dominate under conditions where the bulk flow velocity remains high and capable of advecting photons downward. At
the same time, under conditions typical for accretion columns in X-ray
pulsars, cyclotron processes resulting from thermal collisions must
be treated as part of the bremsstrahlung emissivity and free-free absorption, which reduces their contribution and influences spectral
formation \citep[see][]{Nagel1980, NagelVentura1983, MeszarosNagel1985a}.
This fact, along with the lack of bulk-induced cyclotron emission,
prevents a reliable assessment of the true contribution of cyclotron
emission using the currently available continuum models.

\citetalias{Bissinger2020a} suggested that for conditions where the
cyclotron energy is close to the local plasma temperature, as expected
for \fu's relatively low magnetic field, cyclotron emission from thermal collisions could result in a broad, emission line-like excess. The position of this feature would depend on
the plasma temperature and the cyclotron energy. A similar effect
has been proposed for nonthermal (bulk-induced) collisions
between ambient electrons in the accretion channel and the bulk flow
\citep{Nelson1993}. Simulations by \citet{Nelson1995} showed that this effect can result in the formation of a broad line-like excess in the spectrum. However, the formation of this feature depends on the energy deposition within the emitting region. For bulk-flow collisions, simulations show that collisional excitations mainly occur in the upper (though optically thick) part of the NS atmosphere \citep[see, for example,][]{Miller1989}. Assessing the efficiency and the exact signatures of thermal-motion-induced excitations also requires a simultaneous treatment of the energy balance in the emission region and radiative transfer.

Other possible explanations for the \keVF are based on the fact that spectral formation must include strong polarization effects and radiation redistribution in
a hot plasma, and that the formation
of the continuum and cyclotron lines cannot be fully decoupled. Together, these effects provide a complex continuum
shape characterized by dips and excesses imposed on a general
power-law cutoff \citep{Sokolova2023a}. Depending on the
magnetic field strength and local plasma conditions, several basic
mechanisms for the formation of a \keVF-like hump at intermediate
energies can be outlined. First of all, in a strong magnetic field the two polarization modes
of radiation have principally different energy- and
angle-dependent opacities. 
This behavior is imprinted onto the radiation emerging from a strongly
magnetized medium, leading to characteristic energy-dependent structures in the 
two polarization modes  \citep[see, for example,][]{Nagel1981, MeszarosNagel1985a}, which can, in principle, be visible even in the polarization-averaged
spectrum.
Under typical conditions, the spectrum of the extraordinary mode
generally peaks at intermediate energies, around $\sim$10\,keV, and could therefore 
be responsible for the \keVF \citep{Sokolova2023a}. We note that
the location of the peak varies only mildly with
magnetic field strength \citep{SokolovaThesis}. This scenario for the formation of the
hump could in principle be tested observationally with X-ray polarimetry. 

\begin{figure}
\resizebox{\hsize}{!}{\includegraphics{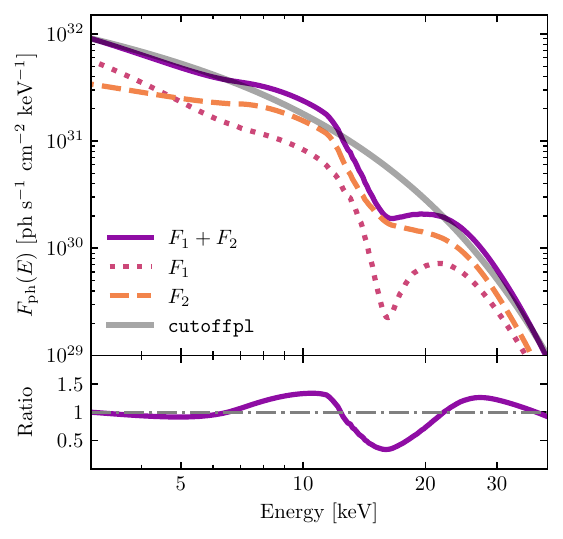}}
  \caption{Top: Modeled photon flux in two polarization modes. The figure emphasizes the formation of a strong red wing due to redistribution near cyclotron resonance. Simulations were performed for a homogeneous slab of self-emitting magnetized plasma with $\tau_\mathrm{T}=1000$, $n_\mathrm{e}=10^{22}\,\mathrm{cm}^{-3}$, $kT_\mathrm{e}=5\,\mathrm{keV}$, and $E_\mathrm{cyc}=20\,\mathrm{keV}$. The spectra for each polarization mode are integrated over angles. Photon polarization modes, mode~1 ($F_1$) and mode~2 ($F_2$), include vacuum effects and plasma polarization \citep[corresponding to the continuous behavior of the refractive indices, see][for more details]{Sokolova2023a}. Assuming a gravitational redshift of $(1+z)=1.24$, the cyclotron line in the final spectrum appears at ${15}\,\mathrm{keV}$. The spectra are shown on the redshifted energy grid. A fit to the power-law spectrum with a high-energy cutoff (\texttt{cutoffpl}), using a photon index of $\Gamma=0.47$, a cutoff energy of $E_\mathrm{fold}=6.6\,\mathrm{keV}$, and a normalization of $\mathrm{N}=2.4\times10^{32}$, is shown in energy flux space to illustrate the generally assumed continuum behavior. Bottom: Ratio between the modeled photon flux and \texttt{cutoffpl}.}
 \label{fig:specsim}
\end{figure}

Another possibility involves a more complex formation of the fundamental line. Since CRSFs arise in spectra due to resonant Compton scattering in a hot plasma, their formation is subject to significant redistribution, which can lead to the development of line wings. At the plasma temperatures typically expected in the accretion channel (4--8\,keV) and under sufficiently strong magnetic fields, it is primarily the red wing of the fundamental line that becomes prominent in the spectra \citep{Alexander1989a, Schoenherr2007a, Schwarm2017b}. Simultaneous treatment of the continuum and cyclotron line formation shows that this broadened red wing can produce a noticeable hump at energies below the fundamental cyclotron resonance \citep{SokolovaThesis}. Interestingly, cyclotron line formation tends to produce a blue wing for moderate magnetic fields when bulk motion Comptonization is considered \citep{Loudas2024a}.

Figure~\ref{fig:specsim} (top) shows an example of spectra in two
polarization modes, obtained from radiative transfer simulations in a
highly magnetized, warm, steady-state plasma. For this exploratory
study, we used the homogeneous model of \citet{MeszarosNagel1985a,MeszarosNagel1985b}, also utilized by \citet{Sokolova2023a},
calculated with the \texttt{FINRAD} transfer code and magnetic
opacities for both the continuum and the fundamental cyclotron resonance. The
transfer was performed in a 1D slab geometry, assuming a
high Thomson optical depth, $\tau_\mathrm{T}=1000$ and an electron
density, $n_\mathrm{e}=10^{22}\,\mathrm{cm}^{-3}$. We set the electron
temperature to $5\,\mathrm{keV}$ and chose a relatively low magnetic field, corresponding to a cyclotron energy of
$20\,\mathrm{keV}$ on the NS surface. 

This setup results in
a cyclotron line at ${\sim}16\,\mathrm{keV}$ in the redshifted,
angle-integrated spectra. The combined spectrum for the
polarization modes show a general power-law-like behavior with a
high-energy cutoff. However, cyclotron resonance,
formed through complex energy and angle redistribution, affects the
spectrum over a wide range of energies. 

We approximate the modeled spectrum as a power law including a high-energy cutoff (\texttt{cutoffpl}) and show the total flux ratio in Fig.~\ref{fig:specsim} (bottom). Below the resonance, the excess in the form of the cyclotron line's red wing is particularly noticeable. Note that in reality, for lower magnetic fields, the
spectral shape of the line's blue wing may be influenced by a second harmonic cyclotron line not included in the theoretical model. 
Although we chose the magnetic field to represent the case of \fu, these simulations are not intended to model the spectrum of this particular source. They merely illustrate the
complexity of the spectra that emerges even within this
simplified framework. These complexities manifest as dip- or hump-like deviations from the power-law continuum. Additional physical
effects, such as the influence of bulk motion and light bending near the NS, may further modify the spectra significantly. 
The deviation of the humps from the continuum, shown in Fig.~\ref{fig:specsim}, is of similar order of magnitude to the relative flux of the \keVF found in our analysis (Figs.~\ref{fig:4u2_residuals}d and~\ref{fig:4u1_residuals}d). It is also compatible with the $\sim$35\% and $\sim$51\% of the continuum flux attributed to the \keVF in the analysis of earlier observations 
by \citetalias{Bissinger2020a}. 
We emphasize that we currently do not fully know how changes of plasma parameters affect changes in the accretion flow. Further effects, such as the influence of temperature in the line-forming region on photon reprocessing, have not yet been fully accounted for. These results are therefore exploratory only.

\section{Summary and conclusions}
In this paper we analyzed short-timescale flux variations in the X-ray spectrum of \fu on timescales of
the system's QPOs and the base-level flux changes. As the flux variations on QPO timescales could be due to changes in the accretion flow, we performed a count-rate-resolved analysis to study the spectral behavior at different intensities. As the individual QPO peaks in the light curve show strong variability (see Fig.~\ref{fig:lc_4u1_zoom}), this selection does not constitute a QPO phase-resolved analysis. After establishing
a model for the
flux-averaged datasets (Sect.~\ref{Sec:SpectralAnalysis}) and then
applying it to the flux-resolved analysis (Sect.~\ref{Sec:Rates}),
we identified a change in the overall spectral shape, particularly in the energy range of the ``10\,keV feature,'' compared to the
flux-averaged data. This suggests the \keVF is an independent spectral component rather than a modeling artifact. 

Assuming an exponentially cutoff power-law continuum, our results provide the strongest evidence to date that the \keVF is likely a separate spectral component.
Specifically, we find that the centroid energy of the \keVF is flux-independent (Fig.~\ref{fig:cm_R}a) within each individual observation, while its relative flux, $F_\mathrm{10keV}/F_\mathrm{PL}$, depends on flux (Fig.~\ref{fig:cm_R}c).
Since flux-dependent changes in the
CRSFs do not affect the \keVF (Fig.~\ref{fig:cm_R}b), these results
strengthen the interpretation of the \keVF as a separate spectral
component and allows us to explain the spectrum without using a second
fundamental CRSF, as sometimes done in the literature. Although both observations show similar principal behavior, 
 $E_\mathrm{10keV}$ differs between them, potentially due to the different source fluxes. If true, this result could suggest some level of coupling between the physical formation of the general continuum and the ``10\,keV feature.'' 

As discussed in Sect.~\ref{Sec:PossibleOrigin}, there are three possible explanations for the ``10\,keV feature.'' First, as speculated by \citetalias{Bissinger2020a}, the \keVF could be due to cyclotron emission. Second, a more complex formation mechanism for the CRSFs, in which Compton scattering produces a broadened red wing of the absorption line, could mimic the feature. 
Finally, polarized radiative transfer that accounts for cyclotron resonance redistribution in strongly magnetized plasmas also yields
a spectral feature in the 10\,keV band \citep{Sokolova2023a}. Further observations, especially with X-ray polarimetry above ${\sim}7\,\mathrm{keV}$ -- that is, at harder energies than those with current missions such as \textit{IXPE} -- would help us choose between these possibilities.

\begin{acknowledgements}
We thank the XMAG collaboration for all the helpful discussions.
This work has been partly funded by DLR under grants 50\,QR\,2103 and 50\,OR\,2410, and by the eROSTEP research unit (WI 1860/17-2).
The material is based on work supported by NASA under award number 80GSFC21M0002 and award number 80GSFC21M0006.  

This research has made use of ISIS functions (ISISscripts) provided by 
ECAP/Remeis observatory and MIT (http://www.sternwarte.uni-erlangen.de/isis/).

This research has made use of data and/or software provided by the High Energy Astrophysics Science Archive Research Center (HEASARC), which is a service of the Astrophysics Science Division at NASA/GSFC. This research has made use of data from the NuSTAR mission, a project led by the California Institute of Technology, managed by the Jet Propulsion Laboratory, and funded by the National Aeronautics and Space Administration. Data analysis was performed using the NuSTAR Data Analysis Software (NuSTARDAS), jointly developed by the ASI Science Data Center (SSDC, Italy) and the California Institute of Technology (USA). We acknowledge the use of public data from the Swift data archive. This work has made use of data from the European Space Agency (ESA) mission \textit{Gaia} (\url{https://www.cosmos.esa.int/gaia}), processed by the \textit{Gaia} Data Processing and Analysis Consortium (DPAC, \url{https://www.cosmos.esa.int/web/gaia/dpac/consortium}). Funding for the DPAC has been provided by national institutions, in particular the institutions participating in the \textit{Gaia} Multilateral Agreement.

\end{acknowledgements}

\bibliographystyle{aa}
\bibliography{references}

@ARTICLE{Alexander1989a,
       author = {{Alexander}, S.~G. and {Meszaros}, P. and {Bussard}, R.~W.},
        title = "{The Nonlinear Transfer Problem in Accreting Pulsars: Stimulated Scattering Effects}",
      journal = {ApJ},
         year = 1989,
        month = jul,
       volume = {342},
        pages = {928},
          doi = {10.1086/167648}
}

@article{Arons1987a,
  title = {Radiation Gasdynamics of Polar {{CAP}} Accretion onto Magnetized Neutron Stars - {{Basic}} Theory},
  author = {Arons, Jonathan and Klein, Richard I. and Lea, Susan M.},
  year = {1987},
  month = jan,
  journal = {ApJ},
  volume = {312},
  pages = {666},
  doi = {10.1086/164912}
}

@ARTICLE{Bachetti2015a,
       author = {{Bachetti}, Matteo and {Harrison}, Fiona A. and {Cook}, Rick and {Tomsick}, John and {Schmid}, Christian and {Grefenstette}, Brian W. and {Barret}, Didier and {Boggs}, Steven E. and {Christensen}, Finn E. and {Craig}, William W. and {Fabian}, Andrew C. and {F{\"u}rst}, Felix and {Gandhi}, Poshak and {Hailey}, Charles J. and {Kara}, Erin and {Maccarone}, Thomas J. and {Miller}, Jon M. and {Pottschmidt}, Katja and {Stern}, Daniel and {Uttley}, Phil and {Walton}, Dominic J. and {Wilms}, J{\"o}rn and {Zhang}, William W.},
        title = "{No Time for Dead Time: Timing Analysis of Bright Black Hole Binaries with NuSTAR}",
      journal = {ApJ},
         year = 2015,
        month = feb,
       volume = {800},
       number = {2},
          eid = {109},
        pages = {109},
          doi = {10.1088/0004-637X/800/2/109},
archivePrefix = {arXiv},
       eprint = {1409.3248},
 primaryClass = {astro-ph.HE}
}

@ARTICLE{Bachetti2018a,
       author = {{Bachetti}, Matteo and {Huppenkothen}, Daniela},
        title = "{No Time for Dead Time: Use the Fourier Amplitude Differences to Normalize Dead-time-affected Periodograms}",
      journal = {ApJ},
         year = 2018,
        month = feb,
       volume = {853},
       number = {2},
          eid = {L21},
        pages = {L21},
          doi = {10.3847/2041-8213/aaa83b},
archivePrefix = {arXiv},
       eprint = {1709.09700},
 primaryClass = {astro-ph.HE}
}

@ARTICLE{Bailer-Jones2021a,
       author = {{Bailer-Jones}, C.~A.~L. and {Rybizki}, J. and {Fouesneau}, M. and {Demleitner}, M. and {Andrae}, R.},
        title = "{Estimating Distances from Parallaxes. V. Geometric and Photogeometric Distances to 1.47 Billion Stars in Gaia Early Data Release 3}",
      journal = {AJ},
         year = 2021,
        month = mar,
       volume = {161},
       number = {3},
          eid = {147},
        pages = {147},
          doi = {10.3847/1538-3881/abd806},
archivePrefix = {arXiv},
       eprint = {2012.05220},
 primaryClass = {astro-ph.SR}
}

@ARTICLE{Barthelmy2005a,
       author = {{Barthelmy}, Scott D. and {Barbier}, Louis M. and {Cummings}, Jay R. and {Fenimore}, Ed E. and {Gehrels}, Neil and {Hullinger}, Derek and {Krimm}, Hans A. and {Markwardt}, Craig B. and {Palmer}, David M. and {Parsons}, Ann and {Sato}, Goro and {Suzuki}, Masaya and {Takahashi}, Tadayuki and {Tashiro}, Makota and {Tueller}, Jack},
        title = "{The Burst Alert Telescope (BAT) on the SWIFT Midex Mission}",
      journal = {Space Sci. Rev.},
         year = 2005,
        month = oct,
       volume = {120},
       number = {3-4},
        pages = {143-164},
          doi = {10.1007/s11214-005-5096-3},
archivePrefix = {arXiv},
       eprint = {astro-ph/0507410},
 primaryClass = {astro-ph}
}

@ARTICLE{Becker2007a,
       author = {{Becker}, Peter A. and {Wolff}, Michael T.},
        title = "{Thermal and Bulk Comptonization in Accretion-powered X-Ray Pulsars}",
      journal = {ApJ},
         year = 2007,
        month = jan,
       volume = {654},
       number = {1},
        pages = {435-457},
          doi = {10.1086/509108},
archivePrefix = {arXiv},
       eprint = {astro-ph/0609035},
 primaryClass = {astro-ph}
}

@article{Becker2022a,
  title = {A {{Generalized Analytical Model}} for {{Thermal}} and {{Bulk Comptonization}} in {{Accretion-powered X-Ray Pulsars}}},
  author = {Becker, Peter A. and Wolff, Michael T.},
  year = {2022},
  month = nov,
  volume = {939},
  pages = {67},
  publisher = {IOP},
  doi = {10.3847/1538-4357/ac8d95},
  journal = {ApJ}
}

@phdthesis{Berger2022a,
    author = {Katrin Berger},
    type = {{MSc} thesis},
    adsurl = {https://www.sternwarte.uni-erlangen.de/docs/theses/2022-10_Berger.pdf},
    school = {Friedrich-Alexander-Universit{\"a}t Erlangen-N{\"u}rnberg},
    title = {Studying the spectral variability of 4U 0115+63},
    year = {2022}
}

@ARTICLE{Bissinger2020a,
       author = {Bissinger, n{\'e} K{\"u}hnel, Matthias and {Kreykenbohm}, Ingo and {Ferrigno}, Carlo and {Pottschmidt}, Katja and {Marcu-Cheatham}, Diana M. and {F{\"u}rst}, Felix and {Rothschild}, Richard E. and {Kretschmar}, Peter and {Klochkov}, Dmitry and {Hemphill}, Paul and {Hertel}, Dominik and {M{\"u}ller}, Sebastian and {Sokolova-Lapa}, Ekaterina and {Oruru}, Bosco and {Grinberg}, Victoria and {Mart{\'\i}nez-N{\'u}{\~n}ez}, Silvia and {Torrej{\'o}n}, Jos{\'e} M. and {Becker}, Peter A. and {Wolff}, Michael T. and {Ballhausen}, Ralf and {Schwarm}, Fritz-Walter and {Wilms}, J{\"o}rn},
        title = "{The giant outburst of 4U 0115+634 in 2011 with Suzaku and RXTE. Minimizing cyclotron line biases}",
      journal = {A\&A},
         year = 2020,
        month = feb,
       volume = {634},
          eid = {A99},
        pages = {A99},
          doi = {10.1051/0004-6361/201935666},
archivePrefix = {arXiv},
       eprint = {1912.06725},
 primaryClass = {astro-ph.HE}
}

@ARTICLE{Burrows2005a,
       author = {{Burrows}, David N. and {Hill}, J.~E. and {Nousek}, J.~A. and {Kennea}, J.~A. and {Wells}, A. and {Osborne}, J.~P. and {Abbey}, A.~F. and {Beardmore}, A. and {Mukerjee}, K. and {Short}, A.~D.~T. and {Chincarini}, G. and {Campana}, S. and {Citterio}, O. and {Moretti}, A. and {Pagani}, C. and {Tagliaferri}, G. and {Giommi}, P. and {Capalbi}, M. and {Tamburelli}, F. and {Angelini}, L. and {Cusumano}, G. and {Br{\"a}uninger}, H.~W. and {Burkert}, W. and {Hartner}, G.~D.},
        title = "{The Swift X-Ray Telescope}",
      journal = {Space Sci. Rev.},
         year = 2005,
        month = oct,
       volume = {120},
       number = {3-4},
        pages = {165-195},
          doi = {10.1007/s11214-005-5097-2},
archivePrefix = {arXiv},
       eprint = {astro-ph/0508071},
 primaryClass = {astro-ph}
}

@ARTICLE{Coburn2002a,
       author = {{Coburn}, W. and {Heindl}, W.~A. and {Rothschild}, R.~E. and {Gruber}, D.~E. and {Kreykenbohm}, I. and {Wilms}, J. and {Kretschmar}, P. and {Staubert}, R.},
        title = "{Magnetic Fields of Accreting X-Ray Pulsars with the Rossi X-Ray Timing Explorer}",
      journal = {ApJ},
         year = 2002,
        month = nov,
       volume = {580},
       number = {1},
        pages = {394-412},
          doi = {10.1086/343033},
archivePrefix = {arXiv},
       eprint = {astro-ph/0207325},
 primaryClass = {astro-ph}
}

@ARTICLE{Cominsky1978a,
       author = {{Cominsky}, L. and {Clark}, G.~W. and {Li}, F. and {Mayer}, W. and {Rappaport}, S.},
        title = "{Discovery of 3.6-s X-ray pulsations from 4U0115+63}",
      journal = {Nature},
         year = 1978,
        month = jun,
       volume = {273},
       number = {5661},
        pages = {367-369},
          doi = {10.1038/273367a0}
}

@ARTICLE{Diez2022a,
       author = {{Diez}, C.~M. and {Grinberg}, V. and {F{\"u}rst}, F. and {Sokolova-Lapa}, E. and {Santangelo}, A. and {Wilms}, J. and {Pottschmidt}, K. and {Mart{\'\i}nez-N{\'u}{\~n}ez}, S. and {Malacaria}, C. and {Kretschmar}, P.},
        title = "{Continuum, cyclotron line, and absorption variability in the high-mass X-ray binary Vela X-1}",
      journal = {A\&A},
         year = 2022,
        month = apr,
       volume = {660},
          eid = {A19},
        pages = {A19},
          doi = {10.1051/0004-6361/202141751},
archivePrefix = {arXiv},
       eprint = {2201.04169},
 primaryClass = {astro-ph.HE}
}

@ARTICLE{Ding2021a,
       author = {{Ding}, Y.~Z. and {Wang}, W. and {Zhang}, P. and {Bu}, Q.~C. and {Cai}, C. and {Cao}, X.~L. and {Zhi}, C. and {Chen}, L. and {Chen}, T.~X. and {Chen}, Y.~B. and {Chen}, Y. and {Chen}, Y.~P. and {Cui}, W.~W. and {Du}, Y.~Y. and {Gao}, G.~H. and {Gao}, H. and {Ge}, M.~Y. and {Gu}, Y.~D. and {Guan}, J. and {Guo}, C.~C. and {Han}, D.~W. and {Huang}, Y. and {Huo}, J. and {Jia}, S.~M. and {Jiang}, W.~C. and {Jin}, J. and {Kong}, L.~D. and {Li}, B. and {Li}, C.~K. and {Li}, G. and {Li}, T.~P. and {Li}, W. and {Li}, X. and {Li}, X.~B. and {Li}, X.~F. and {Li}, Z.~W. and {Liang}, X.~H. and {Liao}, J.~Y. and {Liu}, B.~S. and {Liu}, C.~Z. and {Liu}, H.~X. and {Liu}, H.~W. and {Liu}, X.~J. and {Lu}, F.~J. and {Lu}, X.~F. and {Lou}, Q. and {Luo}, T. and {Ma}, R.~C. and {Ma}, X. and {Meng}, B. and {Nang}, Y. and {Nie}, J.~Y. and {Qu}, J.~L. and {Ren}, X.~Q. and {Sai}, N. and {Song}, L.~M. and {Song}, X.~Y. and {Sun}, L. and {Tan}, Y. and {Tao}, L. and {Tuo}, Y.~L. and {Wang}, C. and {Wang}, L.~J. and {Wang}, P.~J. and {Wang}, W.~S. and {Wang}, Y.~S. and {Wen}, X.~Y. and {Wu}, B.~Y. and {Wu}, B.~B. and {Wu}, M. and {Xiao}, G.~C. and {Xiao}, S. and {Xiong}, S.~L. and {Xu}, Y.~P. and {Yang}, R.~J. and {Yang}, S. and {Yang}, Y.~J. and {Yi}, Q.~B. and {Yin}, Q.~Q. and {You}, Y. and {Zhang}, F. and {Zhang}, H.~M. and {Zhang}, J. and {Zhang}, P. and {Zhang}, S. and {Zhang}, S.~N. and {Zhang}, W.~C. and {Zhang}, W. and {Zhang}, Y.~F. and {Zhang}, Y.~H. and {Zhao}, H.~S. and {Zhao}, X.~F. and {Zheng}, S.~J. and {Zheng}, Y.~G. and {Zhou}, D.~K.},
        title = "{QPOs and orbital elements of X-ray binary 4U 0115+63 during the 2017 outburst observed by Insight-HXMT}",
      journal = {MNRAS},
         year = 2021,
        month = jun,
       volume = {503},
       number = {4},
        pages = {6045-6058},
          doi = {10.1093/mnras/stab835},
archivePrefix = {arXiv},
       eprint = {2102.09498},
 primaryClass = {astro-ph.HE}
}

@ARTICLE{Dugair2013a,
       author = {{Dugair}, Moti R. and {Jaisawal}, Gaurava K. and {Naik}, Sachindra and {Jaaffrey}, S.~N.~A.},
        title = "{Detection of a variable QPO at {\ensuremath{\sim}}41 mHz in the Be/X-ray transient pulsar 4U 0115+634}",
      journal = {MNRAS},
         year = 2013,
        month = sep,
       volume = {434},
       number = {3},
        pages = {2458-2464},
          doi = {10.1093/mnras/stt1187},
archivePrefix = {arXiv},
       eprint = {1308.0919},
 primaryClass = {astro-ph.SR}
}

@ARTICLE{Farinelli2016a,
       author = {{Farinelli}, Ruben and {Ferrigno}, Carlo and {Bozzo}, Enrico and {Becker}, Peter A.},
        title = "{A new model for the X-ray continuum of the magnetized accreting pulsars}",
      journal = {A\&A},
         year = 2016,
        month = jun,
       volume = {591},
          eid = {A29},
        pages = {A29},
          doi = {10.1051/0004-6361/201527257},
archivePrefix = {arXiv},
       eprint = {1602.04308},
 primaryClass = {astro-ph.HE}
}

@ARTICLE{Ferrigno2009a,
       author = {{Ferrigno}, C. and {Becker}, P.~A. and {Segreto}, A. and {Mineo}, T. and {Santangelo}, A.},
        title = "{Study of the accreting pulsar 4U 0115+63 using a bulk and thermal Comptonization model}",
      journal = {A\&A},
         year = 2009,
        month = may,
       volume = {498},
       number = {3},
        pages = {825-836},
          doi = {10.1051/0004-6361/200809373},
archivePrefix = {arXiv},
       eprint = {0902.4392},
 primaryClass = {astro-ph.HE}
}

@ARTICLE{Ferrigno2020a,
       author = {{Ferrigno}, C. and {Bozzo}, E. and {Romano}, P.},
        title = "{Monitoring clumpy wind accretion in supergiant fast-X-ray transients with XMM-Newton}",
      journal = {A\&A},
         year = 2020,
        month = oct,
       volume = {642},
          eid = {A73},
        pages = {A73},
          doi = {10.1051/0004-6361/202038278},
archivePrefix = {arXiv},
       eprint = {2008.04657},
 primaryClass = {astro-ph.HE}
}

@ARTICLE{Fuerst2010a,
       author = {{F{\"u}rst}, F. and {Kreykenbohm}, I. and {Pottschmidt}, K. and {Wilms}, J. and {Hanke}, M. and {Rothschild}, R.~E. and {Kretschmar}, P. and {Schulz}, N.~S. and {Huenemoerder}, D.~P. and {Klochkov}, D. and {Staubert}, R.},
        title = "{X-ray variation statistics and wind clumping in Vela X-1}",
      journal = {A\&A},
         year = 2010,
        month = sep,
       volume = {519},
          eid = {A37},
        pages = {A37},
          doi = {10.1051/0004-6361/200913981},
archivePrefix = {arXiv},
       eprint = {1005.5243},
 primaryClass = {astro-ph.HE}
}

@ARTICLE{Fuerst2014a,
       author = {{F{\"u}rst}, Felix and {Pottschmidt}, Katja and {Wilms}, J{\"o}rn and {Tomsick}, John A. and {Bachetti}, Matteo and {Boggs}, Steven E. and {Christensen}, Finn E. and {Craig}, William W. and {Grefenstette}, Brian W. and {Hailey}, Charles J. and {Harrison}, Fiona and {Madsen}, Kristin K. and {Miller}, Jon M. and {Stern}, Daniel and {Walton}, Dominic J. and {Zhang}, William},
        title = "{NuSTAR Discovery of a Luminosity Dependent Cyclotron Line Energy in Vela X-1}",
      journal = {ApJ},
         year = 2014,
        month = jan,
       volume = {780},
       number = {2},
          eid = {133},
        pages = {133},
          doi = {10.1088/0004-637X/780/2/133},
archivePrefix = {arXiv},
       eprint = {1311.5514},
 primaryClass = {astro-ph.HE}
}

@ARTICLE{GaiaCollab2016a,
       author = {{Gaia Collaboration} and {Prusti}, T. and {de Bruijne}, J.~H.~J. and {Brown}, A.~G.~A. and {Vallenari}, A. and {Babusiaux}, C. and {Bailer-Jones}, C.~A.~L. and {Bastian}, U. and {Biermann}, M. and {Evans}, D.~W. and {Eyer}, L. and {Jansen}, F. and {Jordi}, C. and {Klioner}, S.~A. and {Lammers}, U. and {Lindegren}, L. and {Luri}, X. and {Mignard}, F. and {Milligan}, D.~J. and {Panem}, C. and {Poinsignon}, V. and {Pourbaix}, D. and {Randich}, S. and {Sarri}, G. and {Sartoretti}, P. and {Siddiqui}, H.~I. and {Soubiran}, C. and {Valette}, V. and {van Leeuwen}, F. and {Walton}, N.~A. and {Aerts}, C. and {Arenou}, F. and {Cropper}, M. and {Drimmel}, R. and {H{\o}g}, E. and {Katz}, D. and {Lattanzi}, M.~G. and {O'Mullane}, W. and {Grebel}, E.~K. and {Holland}, A.~D. and {Huc}, C. and {Passot}, X. and {Bramante}, L. and {Cacciari}, C. and {Casta{\~n}eda}, J. and {Chaoul}, L. and {Cheek}, N. and {De Angeli}, F. and {Fabricius}, C. and {Guerra}, R. and {Hern{\'a}ndez}, J. and {Jean-Antoine-Piccolo}, A. and {Masana}, E. and {Messineo}, R. and {Mowlavi}, N. and {Nienartowicz}, K. and {Ord{\'o}{\~n}ez-Blanco}, D. and {Panuzzo}, P. and {Portell}, J. and {Richards}, P.~J. and {Riello}, M. and {Seabroke}, G.~M. and {Tanga}, P. and {Th{\'e}venin}, F. and {Torra}, J. and {Els}, S.~G. and {Gracia-Abril}, G. and {Comoretto}, G. and {Garcia-Reinaldos}, M. and {Lock}, T. and {Mercier}, E. and {Altmann}, M. and {Andrae}, R. and {Astraatmadja}, T.~L. and {Bellas-Velidis}, I. and {Benson}, K. and {Berthier}, J. and {Blomme}, R. and {Busso}, G. and {Carry}, B. and {Cellino}, A. and {Clementini}, G. and {Cowell}, S. and {Creevey}, O. and {Cuypers}, J. and {Davidson}, M. and {De Ridder}, J. and {de Torres}, A. and {Delchambre}, L. and {Dell'Oro}, A. and {Ducourant}, C. and {Fr{\'e}mat}, Y. and {Garc{\'\i}a-Torres}, M. and {Gosset}, E. and {Halbwachs}, J. -L. and {Hambly}, N.~C. and {Harrison}, D.~L. and {Hauser}, M. and {Hestroffer}, D. and {Hodgkin}, S.~T. and {Huckle}, H.~E. and {Hutton}, A. and {Jasniewicz}, G. and {Jordan}, S. and {Kontizas}, M. and {Korn}, A.~J. and {Lanzafame}, A.~C. and {Manteiga}, M. and {Moitinho}, A. and {Muinonen}, K. and {Osinde}, J. and {Pancino}, E. and {Pauwels}, T. and {Petit}, J. -M. and {Recio-Blanco}, A. and {Robin}, A.~C. and {Sarro}, L.~M. and {Siopis}, C. and {Smith}, M. and {Smith}, K.~W. and {Sozzetti}, A. and {Thuillot}, W. and {van Reeven}, W. and {Viala}, Y. and {Abbas}, U. and {Abreu Aramburu}, A. and {Accart}, S. and {Aguado}, J.~J. and {Allan}, P.~M. and {Allasia}, W. and {Altavilla}, G. and {{\'A}lvarez}, M.~A. and {Alves}, J. and {Anderson}, R.~I. and {Andrei}, A.~H. and {Anglada Varela}, E. and {Antiche}, E. and {Antoja}, T. and {Ant{\'o}n}, S. and {Arcay}, B. and {Atzei}, A. and {Ayache}, L. and {Bach}, N. and {Baker}, S.~G. and {Balaguer-N{\'u}{\~n}ez}, L. and {Barache}, C. and {Barata}, C. and {Barbier}, A. and {Barblan}, F. and {Baroni}, M. and {Barrado y Navascu{\'e}s}, D. and {Barros}, M. and {Barstow}, M.~A. and {Becciani}, U. and {Bellazzini}, M. and {Bellei}, G. and {Bello Garc{\'\i}a}, A. and {Belokurov}, V. and {Bendjoya}, P. and {Berihuete}, A. and {Bianchi}, L. and {Bienaym{\'e}}, O. and {Billebaud}, F. and {Blagorodnova}, N. and {Blanco-Cuaresma}, S. and {Boch}, T. and {Bombrun}, A. and {Borrachero}, R. and {Bouquillon}, S. and {Bourda}, G. and {Bouy}, H. and {Bragaglia}, A. and {Breddels}, M.~A. and {Brouillet}, N. and {Br{\"u}semeister}, T. and {Bucciarelli}, B. and {Budnik}, F. and {Burgess}, P. and {Burgon}, R. and {Burlacu}, A. and {Busonero}, D. and {Buzzi}, R. and {Caffau}, E. and {Cambras}, J. and {Campbell}, H. and {Cancelliere}, R. and {Cantat-Gaudin}, T. and {Carlucci}, T. and {Carrasco}, J.~M. and {Castellani}, M. and {Charlot}, P. and {Charnas}, J. and {Charvet}, P. and {Chassat}, F. and {Chiavassa}, A. and {Clotet}, M. and {Cocozza}, G. and {Collins}, R.~S. and {Collins}, P. and {Costigan}, G. and {Crifo}, F. and {Cross}, N.~J.~G. and {Crosta}, M. and {Crowley}, C. and {Dafonte}, C. and {Damerdji}, Y. and {Dapergolas}, A. and {David}, P. and {David}, M. and {De Cat}, P. and {de Felice}, F. and {de Laverny}, P. and {De Luise}, F. and {De March}, R. and {de Martino}, D. and {de Souza}, R. and {Debosscher}, J. and {del Pozo}, E. and {Delbo}, M. and {Delgado}, A. and {Delgado}, H.~E. and {di Marco}, F. and {Di Matteo}, P. and {Diakite}, S. and {Distefano}, E. and {Dolding}, C. and {Dos Anjos}, S. and {Drazinos}, P. and {Dur{\'a}n}, J. and {Dzigan}, Y. and {Ecale}, E. and {Edvardsson}, B. and {Enke}, H. and {Erdmann}, M. and {Escolar}, D. and {Espina}, M. and {Evans}, N.~W. and {Eynard Bontemps}, G. and {Fabre}, C. and {Fabrizio}, M. and {Faigler}, S. and {Falc{\~a}o}, A.~J. and {Farr{\`a}s Casas}, M. and {Faye}, F. and {Federici}, L. and {Fedorets}, G. and {Fern{\'a}ndez-Hern{\'a}ndez}, J. and {Fernique}, P. and {Fienga}, A. and {Figueras}, F. and {Filippi}, F. and {Findeisen}, K. and {Fonti}, A. and {Fouesneau}, M. and {Fraile}, E. and {Fraser}, M. and {Fuchs}, J. and {Furnell}, R. and {Gai}, M. and {Galleti}, S. and {Galluccio}, L. and {Garabato}, D. and {Garc{\'\i}a-Sedano}, F. and {Gar{\'e}}, P. and {Garofalo}, A. and {Garralda}, N. and {Gavras}, P. and {Gerssen}, J. and {Geyer}, R. and {Gilmore}, G. and {Girona}, S. and {Giuffrida}, G. and {Gomes}, M. and {Gonz{\'a}lez-Marcos}, A. and {Gonz{\'a}lez-N{\'u}{\~n}ez}, J. and {Gonz{\'a}lez-Vidal}, J.~J. and {Granvik}, M. and {Guerrier}, A. and {Guillout}, P. and {Guiraud}, J. and {G{\'u}rpide}, A. and {Guti{\'e}rrez-S{\'a}nchez}, R. and {Guy}, L.~P. and {Haigron}, R. and {Hatzidimitriou}, D. and {Haywood}, M. and {Heiter}, U. and {Helmi}, A. and {Hobbs}, D. and {Hofmann}, W. and {Holl}, B. and {Holland}, G. and {Hunt}, J.~A.~S. and {Hypki}, A. and {Icardi}, V. and {Irwin}, M. and {Jevardat de Fombelle}, G. and {Jofr{\'e}}, P. and {Jonker}, P.~G. and {Jorissen}, A. and {Julbe}, F. and {Karampelas}, A. and {Kochoska}, A. and {Kohley}, R. and {Kolenberg}, K. and {Kontizas}, E. and {Koposov}, S.~E. and {Kordopatis}, G. and {Koubsky}, P. and {Kowalczyk}, A. and {Krone-Martins}, A. and {Kudryashova}, M. and {Kull}, I. and {Bachchan}, R.~K. and {Lacoste-Seris}, F. and {Lanza}, A.~F. and {Lavigne}, J. -B. and {Le Poncin-Lafitte}, C. and {Lebreton}, Y. and {Lebzelter}, T. and {Leccia}, S. and {Leclerc}, N. and {Lecoeur-Taibi}, I. and {Lemaitre}, V. and {Lenhardt}, H. and {Leroux}, F. and {Liao}, S. and {Licata}, E. and {Lindstr{\o}m}, H.~E.~P. and {Lister}, T.~A. and {Livanou}, E. and {Lobel}, A. and {L{\"o}ffler}, W. and {L{\'o}pez}, M. and {Lopez-Lozano}, A. and {Lorenz}, D. and {Loureiro}, T. and {MacDonald}, I. and {Magalh{\~a}es Fernandes}, T. and {Managau}, S. and {Mann}, R.~G. and {Mantelet}, G. and {Marchal}, O. and {Marchant}, J.~M. and {Marconi}, M. and {Marie}, J. and {Marinoni}, S. and {Marrese}, P.~M. and {Marschalk{\'o}}, G. and {Marshall}, D.~J. and {Mart{\'\i}n-Fleitas}, J.~M. and {Martino}, M. and {Mary}, N. and {Matijevi{\v{c}}}, G. and {Mazeh}, T. and {McMillan}, P.~J. and {Messina}, S. and {Mestre}, A. and {Michalik}, D. and {Millar}, N.~R. and {Miranda}, B.~M.~H. and {Molina}, D. and {Molinaro}, R. and {Molinaro}, M. and {Moln{\'a}r}, L. and {Moniez}, M. and {Montegriffo}, P. and {Monteiro}, D. and {Mor}, R. and {Mora}, A. and {Morbidelli}, R. and {Morel}, T. and {Morgenthaler}, S. and {Morley}, T. and {Morris}, D. and {Mulone}, A.~F. and {Muraveva}, T. and {Musella}, I. and {Narbonne}, J. and {Nelemans}, G. and {Nicastro}, L. and {Noval}, L. and {Ord{\'e}novic}, C. and {Ordieres-Mer{\'e}}, J. and {Osborne}, P. and {Pagani}, C. and {Pagano}, I. and {Pailler}, F. and {Palacin}, H. and {Palaversa}, L. and {Parsons}, P. and {Paulsen}, T. and {Pecoraro}, M. and {Pedrosa}, R. and {Pentik{\"a}inen}, H. and {Pereira}, J. and {Pichon}, B. and {Piersimoni}, A.~M. and {Pineau}, F. -X. and {Plachy}, E. and {Plum}, G. and {Poujoulet}, E. and {Pr{\v{s}}a}, A. and {Pulone}, L. and {Ragaini}, S. and {Rago}, S. and {Rambaux}, N. and {Ramos-Lerate}, M. and {Ranalli}, P. and {Rauw}, G. and {Read}, A. and {Regibo}, S. and {Renk}, F. and {Reyl{\'e}}, C. and {Ribeiro}, R.~A. and {Rimoldini}, L. and {Ripepi}, V. and {Riva}, A. and {Rixon}, G. and {Roelens}, M. and {Romero-G{\'o}mez}, M. and {Rowell}, N. and {Royer}, F. and {Rudolph}, A. and {Ruiz-Dern}, L. and {Sadowski}, G. and {Sagrist{\`a} Sell{\'e}s}, T. and {Sahlmann}, J. and {Salgado}, J. and {Salguero}, E. and {Sarasso}, M. and {Savietto}, H. and {Schnorhk}, A. and {Schultheis}, M. and {Sciacca}, E. and {Segol}, M. and {Segovia}, J.~C. and {Segransan}, D. and {Serpell}, E. and {Shih}, I. -C. and {Smareglia}, R. and {Smart}, R.~L. and {Smith}, C. and {Solano}, E. and {Solitro}, F. and {Sordo}, R. and {Soria Nieto}, S. and {Souchay}, J. and {Spagna}, A. and {Spoto}, F. and {Stampa}, U. and {Steele}, I.~A. and {Steidelm{\"u}ller}, H. and {Stephenson}, C.~A. and {Stoev}, H. and {Suess}, F.~F. and {S{\"u}veges}, M. and {Surdej}, J. and {Szabados}, L. and {Szegedi-Elek}, E. and {Tapiador}, D. and {Taris}, F. and {Tauran}, G. and {Taylor}, M.~B. and {Teixeira}, R. and {Terrett}, D. and {Tingley}, B. and {Trager}, S.~C. and {Turon}, C. and {Ulla}, A. and {Utrilla}, E. and {Valentini}, G. and {van Elteren}, A. and {Van Hemelryck}, E. and {van Leeuwen}, M. and {Varadi}, M. and {Vecchiato}, A. and {Veljanoski}, J. and {Via}, T. and {Vicente}, D. and {Vogt}, S. and {Voss}, H. and {Votruba}, V. and {Voutsinas}, S. and {Walmsley}, G. and {Weiler}, M. and {Weingrill}, K. and {Werner}, D. and {Wevers}, T. and {Whitehead}, G. and {Wyrzykowski}, {\L}. and {Yoldas}, A. and {{\v{Z}}erjal}, M. and {Zucker}, S. and {Zurbach}, C. and {Zwitter}, T. and {Alecu}, A. and {Allen}, M. and {Allende Prieto}, C. and {Amorim}, A. and {Anglada-Escud{\'e}}, G. and {Arsenijevic}, V. and {Azaz}, S. and {Balm}, P. and {Beck}, M. and {Bernstein}, H. -H. and {Bigot}, L. and {Bijaoui}, A. and {Blasco}, C. and {Bonfigli}, M. and {Bono}, G. and {Boudreault}, S. and {Bressan}, A. and {Brown}, S. and {Brunet}, P. -M. and {Bunclark}, P. and {Buonanno}, R. and {Butkevich}, A.~G. and {Carret}, C. and {Carrion}, C. and {Chemin}, L. and {Ch{\'e}reau}, F. and {Corcione}, L. and {Darmigny}, E. and {de Boer}, K.~S. and {de Teodoro}, P. and {de Zeeuw}, P.~T. and {Delle Luche}, C. and {Domingues}, C.~D. and {Dubath}, P. and {Fodor}, F. and {Fr{\'e}zouls}, B. and {Fries}, A. and {Fustes}, D. and {Fyfe}, D. and {Gallardo}, E. and {Gallegos}, J. and {Gardiol}, D. and {Gebran}, M. and {Gomboc}, A. and {G{\'o}mez}, A. and {Grux}, E. and {Gueguen}, A. and {Heyrovsky}, A. and {Hoar}, J. and {Iannicola}, G. and {Isasi Parache}, Y. and {Janotto}, A. -M. and {Joliet}, E. and {Jonckheere}, A. and {Keil}, R. and {Kim}, D. -W. and {Klagyivik}, P. and {Klar}, J. and {Knude}, J. and {Kochukhov}, O. and {Kolka}, I. and {Kos}, J. and {Kutka}, A. and {Lainey}, V. and {LeBouquin}, D. and {Liu}, C. and {Loreggia}, D. and {Makarov}, V.~V. and {Marseille}, M.~G. and {Martayan}, C. and {Martinez-Rubi}, O. and {Massart}, B. and {Meynadier}, F. and {Mignot}, S. and {Munari}, U. and {Nguyen}, A. -T. and {Nordlander}, T. and {Ocvirk}, P. and {O'Flaherty}, K.~S. and {Olias Sanz}, A. and {Ortiz}, P. and {Osorio}, J. and {Oszkiewicz}, D. and {Ouzounis}, A. and {Palmer}, M. and {Park}, P. and {Pasquato}, E. and {Peltzer}, C. and {Peralta}, J. and {P{\'e}turaud}, F. and {Pieniluoma}, T. and {Pigozzi}, E. and {Poels}, J. and {Prat}, G. and {Prod'homme}, T. and {Raison}, F. and {Rebordao}, J.~M. and {Risquez}, D. and {Rocca-Volmerange}, B. and {Rosen}, S. and {Ruiz-Fuertes}, M.~I. and {Russo}, F. and {Sembay}, S. and {Serraller Vizcaino}, I. and {Short}, A. and {Siebert}, A. and {Silva}, H. and {Sinachopoulos}, D. and {Slezak}, E. and {Soffel}, M. and {Sosnowska}, D. and {Strai{\v{z}}ys}, V. and {ter Linden}, M. and {Terrell}, D. and {Theil}, S. and {Tiede}, C. and {Troisi}, L. and {Tsalmantza}, P. and {Tur}, D. and {Vaccari}, M. and {Vachier}, F. and {Valles}, P. and {Van Hamme}, W. and {Veltz}, L. and {Virtanen}, J. and {Wallut}, J. -M. and {Wichmann}, R. and {Wilkinson}, M.~I. and {Ziaeepour}, H. and {Zschocke}, S.},
        title = "{The Gaia mission}",
      journal = {A\&A},
         year = 2016,
        month = nov,
       volume = {595},
          eid = {A1},
        pages = {A1},
          doi = {10.1051/0004-6361/201629272},
archivePrefix = {arXiv},
       eprint = {1609.04153},
 primaryClass = {astro-ph.IM}
}

@ARTICLE{GaiaCollab2023a,
       author = {{Gaia Collaboration} and {Vallenari}, A. and {Brown}, A.~G.~A. and {Prusti}, T. and {de Bruijne}, J.~H.~J. and {Arenou}, F. and {Babusiaux}, C. and {Biermann}, M. and {Creevey}, O.~L. and {Ducourant}, C. and {Evans}, D.~W. and {Eyer}, L. and {Guerra}, R. and {Hutton}, A. and {Jordi}, C. and {Klioner}, S.~A. and {Lammers}, U.~L. and {Lindegren}, L. and {Luri}, X. and {Mignard}, F. and {Panem}, C. and {Pourbaix}, D. and {Randich}, S. and {Sartoretti}, P. and {Soubiran}, C. and {Tanga}, P. and {Walton}, N.~A. and {Bailer-Jones}, C.~A.~L. and {Bastian}, U. and {Drimmel}, R. and {Jansen}, F. and {Katz}, D. and {Lattanzi}, M.~G. and {van Leeuwen}, F. and {Bakker}, J. and {Cacciari}, C. and {Casta{\~n}eda}, J. and {De Angeli}, F. and {Fabricius}, C. and {Fouesneau}, M. and {Fr{\'e}mat}, Y. and {Galluccio}, L. and {Guerrier}, A. and {Heiter}, U. and {Masana}, E. and {Messineo}, R. and {Mowlavi}, N. and {Nicolas}, C. and {Nienartowicz}, K. and {Pailler}, F. and {Panuzzo}, P. and {Riclet}, F. and {Roux}, W. and {Seabroke}, G.~M. and {Sordo}, R. and {Th{\'e}venin}, F. and {Gracia-Abril}, G. and {Portell}, J. and {Teyssier}, D. and {Altmann}, M. and {Andrae}, R. and {Audard}, M. and {Bellas-Velidis}, I. and {Benson}, K. and {Berthier}, J. and {Blomme}, R. and {Burgess}, P.~W. and {Busonero}, D. and {Busso}, G. and {C{\'a}novas}, H. and {Carry}, B. and {Cellino}, A. and {Cheek}, N. and {Clementini}, G. and {Damerdji}, Y. and {Davidson}, M. and {de Teodoro}, P. and {Nu{\~n}ez Campos}, M. and {Delchambre}, L. and {Dell'Oro}, A. and {Esquej}, P. and {Fern{\'a}ndez-Hern{\'a}ndez}, J. and {Fraile}, E. and {Garabato}, D. and {Garc{\'\i}a-Lario}, P. and {Gosset}, E. and {Haigron}, R. and {Halbwachs}, J. -L. and {Hambly}, N.~C. and {Harrison}, D.~L. and {Hern{\'a}ndez}, J. and {Hestroffer}, D. and {Hodgkin}, S.~T. and {Holl}, B. and {Jan{\ss}en}, K. and {Jevardat de Fombelle}, G. and {Jordan}, S. and {Krone-Martins}, A. and {Lanzafame}, A.~C. and {L{\"o}ffler}, W. and {Marchal}, O. and {Marrese}, P.~M. and {Moitinho}, A. and {Muinonen}, K. and {Osborne}, P. and {Pancino}, E. and {Pauwels}, T. and {Recio-Blanco}, A. and {Reyl{\'e}}, C. and {Riello}, M. and {Rimoldini}, L. and {Roegiers}, T. and {Rybizki}, J. and {Sarro}, L.~M. and {Siopis}, C. and {Smith}, M. and {Sozzetti}, A. and {Utrilla}, E. and {van Leeuwen}, M. and {Abbas}, U. and {{\'A}brah{\'a}m}, P. and {Abreu Aramburu}, A. and {Aerts}, C. and {Aguado}, J.~J. and {Ajaj}, M. and {Aldea-Montero}, F. and {Altavilla}, G. and {{\'A}lvarez}, M.~A. and {Alves}, J. and {Anders}, F. and {Anderson}, R.~I. and {Anglada Varela}, E. and {Antoja}, T. and {Baines}, D. and {Baker}, S.~G. and {Balaguer-N{\'u}{\~n}ez}, L. and {Balbinot}, E. and {Balog}, Z. and {Barache}, C. and {Barbato}, D. and {Barros}, M. and {Barstow}, M.~A. and {Bartolom{\'e}}, S. and {Bassilana}, J. -L. and {Bauchet}, N. and {Becciani}, U. and {Bellazzini}, M. and {Berihuete}, A. and {Bernet}, M. and {Bertone}, S. and {Bianchi}, L. and {Binnenfeld}, A. and {Blanco-Cuaresma}, S. and {Blazere}, A. and {Boch}, T. and {Bombrun}, A. and {Bossini}, D. and {Bouquillon}, S. and {Bragaglia}, A. and {Bramante}, L. and {Breedt}, E. and {Bressan}, A. and {Brouillet}, N. and {Brugaletta}, E. and {Bucciarelli}, B. and {Burlacu}, A. and {Butkevich}, A.~G. and {Buzzi}, R. and {Caffau}, E. and {Cancelliere}, R. and {Cantat-Gaudin}, T. and {Carballo}, R. and {Carlucci}, T. and {Carnerero}, M.~I. and {Carrasco}, J.~M. and {Casamiquela}, L. and {Castellani}, M. and {Castro-Ginard}, A. and {Chaoul}, L. and {Charlot}, P. and {Chemin}, L. and {Chiaramida}, V. and {Chiavassa}, A. and {Chornay}, N. and {Comoretto}, G. and {Contursi}, G. and {Cooper}, W.~J. and {Cornez}, T. and {Cowell}, S. and {Crifo}, F. and {Cropper}, M. and {Crosta}, M. and {Crowley}, C. and {Dafonte}, C. and {Dapergolas}, A. and {David}, M. and {David}, P. and {de Laverny}, P. and {De Luise}, F. and {De March}, R. and {De Ridder}, J. and {de Souza}, R. and {de Torres}, A. and {del Peloso}, E.~F. and {del Pozo}, E. and {Delbo}, M. and {Delgado}, A. and {Delisle}, J. -B. and {Demouchy}, C. and {Dharmawardena}, T.~E. and {Di Matteo}, P. and {Diakite}, S. and {Diener}, C. and {Distefano}, E. and {Dolding}, C. and {Edvardsson}, B. and {Enke}, H. and {Fabre}, C. and {Fabrizio}, M. and {Faigler}, S. and {Fedorets}, G. and {Fernique}, P. and {Fienga}, A. and {Figueras}, F. and {Fournier}, Y. and {Fouron}, C. and {Fragkoudi}, F. and {Gai}, M. and {Garcia-Gutierrez}, A. and {Garcia-Reinaldos}, M. and {Garc{\'\i}a-Torres}, M. and {Garofalo}, A. and {Gavel}, A. and {Gavras}, P. and {Gerlach}, E. and {Geyer}, R. and {Giacobbe}, P. and {Gilmore}, G. and {Girona}, S. and {Giuffrida}, G. and {Gomel}, R. and {Gomez}, A. and {Gonz{\'a}lez-N{\'u}{\~n}ez}, J. and {Gonz{\'a}lez-Santamar{\'\i}a}, I. and {Gonz{\'a}lez-Vidal}, J.~J. and {Granvik}, M. and {Guillout}, P. and {Guiraud}, J. and {Guti{\'e}rrez-S{\'a}nchez}, R. and {Guy}, L.~P. and {Hatzidimitriou}, D. and {Hauser}, M. and {Haywood}, M. and {Helmer}, A. and {Helmi}, A. and {Sarmiento}, M.~H. and {Hidalgo}, S.~L. and {Hilger}, T. and {H{\l}adczuk}, N. and {Hobbs}, D. and {Holland}, G. and {Huckle}, H.~E. and {Jardine}, K. and {Jasniewicz}, G. and {Jean-Antoine Piccolo}, A. and {Jim{\'e}nez-Arranz}, {\'O}. and {Jorissen}, A. and {Juaristi Campillo}, J. and {Julbe}, F. and {Karbevska}, L. and {Kervella}, P. and {Khanna}, S. and {Kontizas}, M. and {Kordopatis}, G. and {Korn}, A.~J. and {K{\'o}sp{\'a}l}, {\'A}. and {Kostrzewa-Rutkowska}, Z. and {Kruszy{\'n}ska}, K. and {Kun}, M. and {Laizeau}, P. and {Lambert}, S. and {Lanza}, A.~F. and {Lasne}, Y. and {Le Campion}, J. -F. and {Lebreton}, Y. and {Lebzelter}, T. and {Leccia}, S. and {Leclerc}, N. and {Lecoeur-Taibi}, I. and {Liao}, S. and {Licata}, E.~L. and {Lindstr{\o}m}, H.~E.~P. and {Lister}, T.~A. and {Livanou}, E. and {Lobel}, A. and {Lorca}, A. and {Loup}, C. and {Madrero Pardo}, P. and {Magdaleno Romeo}, A. and {Managau}, S. and {Mann}, R.~G. and {Manteiga}, M. and {Marchant}, J.~M. and {Marconi}, M. and {Marcos}, J. and {Marcos Santos}, M.~M.~S. and {Mar{\'\i}n Pina}, D. and {Marinoni}, S. and {Marocco}, F. and {Marshall}, D.~J. and {Martin Polo}, L. and {Mart{\'\i}n-Fleitas}, J.~M. and {Marton}, G. and {Mary}, N. and {Masip}, A. and {Massari}, D. and {Mastrobuono-Battisti}, A. and {Mazeh}, T. and {McMillan}, P.~J. and {Messina}, S. and {Michalik}, D. and {Millar}, N.~R. and {Mints}, A. and {Molina}, D. and {Molinaro}, R. and {Moln{\'a}r}, L. and {Monari}, G. and {Mongui{\'o}}, M. and {Montegriffo}, P. and {Montero}, A. and {Mor}, R. and {Mora}, A. and {Morbidelli}, R. and {Morel}, T. and {Morris}, D. and {Muraveva}, T. and {Murphy}, C.~P. and {Musella}, I. and {Nagy}, Z. and {Noval}, L. and {Oca{\~n}a}, F. and {Ogden}, A. and {Ordenovic}, C. and {Osinde}, J.~O. and {Pagani}, C. and {Pagano}, I. and {Palaversa}, L. and {Palicio}, P.~A. and {Pallas-Quintela}, L. and {Panahi}, A. and {Payne-Wardenaar}, S. and {Pe{\~n}alosa Esteller}, X. and {Penttil{\"a}}, A. and {Pichon}, B. and {Piersimoni}, A.~M. and {Pineau}, F. -X. and {Plachy}, E. and {Plum}, G. and {Poggio}, E. and {Pr{\v{s}}a}, A. and {Pulone}, L. and {Racero}, E. and {Ragaini}, S. and {Rainer}, M. and {Raiteri}, C.~M. and {Rambaux}, N. and {Ramos}, P. and {Ramos-Lerate}, M. and {Re Fiorentin}, P. and {Regibo}, S. and {Richards}, P.~J. and {Rios Diaz}, C. and {Ripepi}, V. and {Riva}, A. and {Rix}, H. -W. and {Rixon}, G. and {Robichon}, N. and {Robin}, A.~C. and {Robin}, C. and {Roelens}, M. and {Rogues}, H.~R.~O. and {Rohrbasser}, L. and {Romero-G{\'o}mez}, M. and {Rowell}, N. and {Royer}, F. and {Ruz Mieres}, D. and {Rybicki}, K.~A. and {Sadowski}, G. and {S{\'a}ez N{\'u}{\~n}ez}, A. and {Sagrist{\`a} Sell{\'e}s}, A. and {Sahlmann}, J. and {Salguero}, E. and {Samaras}, N. and {Sanchez Gimenez}, V. and {Sanna}, N. and {Santove{\~n}a}, R. and {Sarasso}, M. and {Schultheis}, M. and {Sciacca}, E. and {Segol}, M. and {Segovia}, J.~C. and {S{\'e}gransan}, D. and {Semeux}, D. and {Shahaf}, S. and {Siddiqui}, H.~I. and {Siebert}, A. and {Siltala}, L. and {Silvelo}, A. and {Slezak}, E. and {Slezak}, I. and {Smart}, R.~L. and {Snaith}, O.~N. and {Solano}, E. and {Solitro}, F. and {Souami}, D. and {Souchay}, J. and {Spagna}, A. and {Spina}, L. and {Spoto}, F. and {Steele}, I.~A. and {Steidelm{\"u}ller}, H. and {Stephenson}, C.~A. and {S{\"u}veges}, M. and {Surdej}, J. and {Szabados}, L. and {Szegedi-Elek}, E. and {Taris}, F. and {Taylor}, M.~B. and {Teixeira}, R. and {Tolomei}, L. and {Tonello}, N. and {Torra}, F. and {Torra}, J. and {Torralba Elipe}, G. and {Trabucchi}, M. and {Tsounis}, A.~T. and {Turon}, C. and {Ulla}, A. and {Unger}, N. and {Vaillant}, M.~V. and {van Dillen}, E. and {van Reeven}, W. and {Vanel}, O. and {Vecchiato}, A. and {Viala}, Y. and {Vicente}, D. and {Voutsinas}, S. and {Weiler}, M. and {Wevers}, T. and {Wyrzykowski}, {\L}. and {Yoldas}, A. and {Yvard}, P. and {Zhao}, H. and {Zorec}, J. and {Zucker}, S. and {Zwitter}, T.},
        title = "{Gaia Data Release 3. Summary of the content and survey properties}",
      journal = {A\&A},
         year = 2023,
        month = jun,
       volume = {674},
          eid = {A1},
        pages = {A1},
          doi = {10.1051/0004-6361/202243940},
archivePrefix = {arXiv},
       eprint = {2208.00211},
 primaryClass = {astro-ph.GA}
}

@ARTICLE{Gehrels2004a,
       author = {{Gehrels}, N. and {Swift Team}},
        title = "{The Swift {\ensuremath{\gamma}}-ray burst mission}",
      journal = {New A Rev.},
         year = 2004,
        month = apr,
       volume = {48},
       number = {5-6},
        pages = {431-435},
          doi = {10.1016/j.newar.2003.12.055}
}

@ARTICLE{Giacconi1972a,
       author = {{Giacconi}, R. and {Murray}, S. and {Gursky}, H. and {Kellogg}, E. and {Schreier}, E. and {Tananbaum}, H.},
        title = "{The Uhuru catalog of X-ray sources.}",
      journal = {ApJ},
         year = 1972,
        month = dec,
       volume = {178},
        pages = {281-308},
          doi = {10.1086/151790}
}

@ARTICLE{Harrison2013a,
       author = {{Harrison}, Fiona A. and {Craig}, William W. and {Christensen}, Finn E. and {Hailey}, Charles J. and {Zhang}, William W. and {Boggs}, Steven E. and {Stern}, Daniel and {Cook}, W. Rick and {Forster}, Karl and {Giommi}, Paolo and {Grefenstette}, Brian W. and {Kim}, Yunjin and {Kitaguchi}, Takao and {Koglin}, Jason E. and {Madsen}, Kristin K. and {Mao}, Peter H. and {Miyasaka}, Hiromasa and {Mori}, Kaya and {Perri}, Matteo and {Pivovaroff}, Michael J. and {Puccetti}, Simonetta and {Rana}, Vikram R. and {Westergaard}, Niels J. and {Willis}, Jason and {Zoglauer}, Andreas and {An}, Hongjun and {Bachetti}, Matteo and {Barri{\`e}re}, Nicolas M. and {Bellm}, Eric C. and {Bhalerao}, Varun and {Brejnholt}, Nicolai F. and {Fuerst}, Felix and {Liebe}, Carl C. and {Markwardt}, Craig B. and {Nynka}, Melania and {Vogel}, Julia K. and {Walton}, Dominic J. and {Wik}, Daniel R. and {Alexander}, David M. and {Cominsky}, Lynn R. and {Hornschemeier}, Ann E. and {Hornstrup}, Allan and {Kaspi}, Victoria M. and {Madejski}, Greg M. and {Matt}, Giorgio and {Molendi}, Silvano and {Smith}, David M. and {Tomsick}, John A. and {Ajello}, Marco and {Ballantyne}, David R. and {Balokovi{\'c}}, Mislav and {Barret}, Didier and {Bauer}, Franz E. and {Blandford}, Roger D. and {Brandt}, W. Niel and {Brenneman}, Laura W. and {Chiang}, James and {Chakrabarty}, Deepto and {Chenevez}, Jerome and {Comastri}, Andrea and {Dufour}, Francois and {Elvis}, Martin and {Fabian}, Andrew C. and {Farrah}, Duncan and {Fryer}, Chris L. and {Gotthelf}, Eric V. and {Grindlay}, Jonathan E. and {Helfand}, David J. and {Krivonos}, Roman and {Meier}, David L. and {Miller}, Jon M. and {Natalucci}, Lorenzo and {Ogle}, Patrick and {Ofek}, Eran O. and {Ptak}, Andrew and {Reynolds}, Stephen P. and {Rigby}, Jane R. and {Tagliaferri}, Gianpiero and {Thorsett}, Stephen E. and {Treister}, Ezequiel and {Urry}, C. Megan},
        title = "{The Nuclear Spectroscopic Telescope Array (NuSTAR) High-energy X-Ray Mission}",
      journal = {ApJ},
         year = 2013,
        month = jun,
       volume = {770},
       number = {2},
          eid = {103},
        pages = {103},
          doi = {10.1088/0004-637X/770/2/103},
archivePrefix = {arXiv},
       eprint = {1301.7307},
 primaryClass = {astro-ph.IM}
}

@ARTICLE{Heindl1999a,
       author = {{Heindl}, W.~A. and {Coburn}, W. and {Gruber}, D.~E. and {Pelling}, M.~R. and {Rothschild}, R.~E. and {Wilms}, J. and {Pottschmidt}, K. and {Staubert}, R.},
        title = "{Discovery of a Third Harmonic Cyclotron Resonance Scattering Feature in the X-Ray Spectrum of 4U 0115+63}",
      journal = {ApJ},
         year = 1999,
        month = aug,
       volume = {521},
       number = {1},
        pages = {L49-L53},
          doi = {10.1086/312172},
archivePrefix = {arXiv},
       eprint = {astro-ph/9904222},
 primaryClass = {astro-ph}
}

@article{Heindl2000a,
       author = {{Heindl}, W.~A. and {Coburn}, W. and {Gruber}, D.~E. and {Pelling}, M.~R. and {Rothschild}, R.~E. and {Kretschmar}, P. and {Kreykenbohm}, I. and {Pottschmidt}, K. and {Staubert}, R. and {Wilms}, J.},
        title = "{Multiple Cyclotron Lines in the Spectrum of 4U 0115+63}",
       journal = {BAAS},
         year = 2000,
       series = {AAS/High Energy Astrophysics Division},
volume={32},
month = oct,
        pages = {1230}
}

@INPROCEEDINGS{Houck2000a,
       author = {{Houck}, J.~C. and {Denicola}, L.~A.},
        title = "{ISIS: An Interactive Spectral Interpretation System for High Resolution X-Ray Spectroscopy}",
    booktitle = {Astronomical Data Analysis Software and Systems IX},
         year = 2000,
       editor = {{Manset}, Nadine and {Veillet}, Christian and {Crabtree}, Dennis},
       series = {Astronomical Society of the Pacific Conference Series},
       volume = {216},
        month = jan,
        pages = {591}
}

@ARTICLE{Iyer2015a,
       author = {{Iyer}, N. and {Mukherjee}, D. and {Dewangan}, G.~C. and {Bhattacharya}, D. and {Seetha}, S.},
        title = "{Variations in the cyclotron resonant scattering features during 2011 outburst of 4U 0115+63}",
      journal = {MNRAS},
         year = 2015,
        month = nov,
       volume = {454},
       number = {1},
        pages = {741-751},
          doi = {10.1093/mnras/stv1942},
archivePrefix = {arXiv},
       eprint = {1506.03376},
 primaryClass = {astro-ph.HE}
}

@ARTICLE{Jain2025a,
       author = {{Jain}, Chetana and {Sharma}, Prince and {Dutta}, Anjan},
        title = "{Timing analysis of Be/X-ray transient 4U 0115 + 63 during Type II outburst of 2023 using NuSTAR observations}",
      journal = {Adv.\ Space Res.},
         year = 2025,
        month = jan,
       volume = {75},
       number = {1},
        pages = {1490-1501},
          doi = {10.1016/j.asr.2024.10.021}
}

@ARTICLE{Lampton1976a,
       author = {{Lampton}, M. and {Margon}, B. and {Bowyer}, S.},
        title = "{Parameter estimation in X-ray astronomy.}",
      journal = {ApJ},
         year = 1976,
        month = aug,
       volume = {208},
        pages = {177-190},
          doi = {10.1086/154592}
}

@ARTICLE{Li2012a,
       author = {{Li}, Jun and {Wang}, Wei and {Zhao}, Yongheng},
        title = "{Cyclotron resonance energies and orbital elements of accretion pulsar 4U 0115+63 during the giant outburst in 2008}",
      journal = {MNRAS},
         year = 2012,
        month = jul,
       volume = {423},
       number = {3},
        pages = {2854-2867},
          doi = {10.1111/j.1365-2966.2012.21096.x},
       eprint = {1204.2908},
 primaryClass = {astro-ph.HE}
}

@ARTICLE{Li2024a,
       author = {{Li}, Panping P. and {Becker}, Peter A. and {Tao}, Lian},
        title = "{Spectral evolution of RX J0440.9+4431 during the 2022{\textendash}2023 giant outburst observed with Insight-HXMT}",
      journal = {A\&A},
         year = 2024,
        month = sep,
       volume = {689},
          eid = {A316},
        pages = {A316},
          doi = {10.1051/0004-6361/202450149},
       eprint = {2407.01586},
 primaryClass = {astro-ph.HE}
}

@ARTICLE{Li2025a,
       author = {{Li}, P.~P. and {Tao}, L. and {Ma}, R.~C. and {Zhao}, Q.~C. and {Zhang}, L. and {Zhao}, S.~J. and {Huang}, Y. and {Ma}, X. and {Feng}, H. and {Li}, Z.~X. and {Yang}, Z.~H.},
        title = "{Study of millihertz quasiperiodic oscillations in the 2023 outburst of 4U 0115+63 using NuSTAR}",
      journal = {A\&A},
         year = 2025,
        month = nov,
       volume = {703},
          eid = {A124},
        pages = {A124},
          doi = {10.1051/0004-6361/202555558}
}

@ARTICLE{Liu2020a,
       author = {{Liu}, Bai-Sheng and {Tao}, Lian and {Zhang}, Shuang-Nan and {Li}, Xiang-Dong and {Ge}, Ming-Yu and {Qu}, Jin-Lu and {Song}, Li-Ming and {Ji}, Long and {Zhang}, Shu and {Santangelo}, Andrea and {Wang}, Ling-Jun},
        title = "{A Peculiar Cyclotron Line near 16 keV Detected in the 2015 Outburst of 4U 0115+63?}",
      journal = {ApJ},
         year = 2020,
        month = sep,
       volume = {900},
       number = {1},
          eid = {41},
        pages = {41},
          doi = {10.3847/1538-4357/aba4a5},
archivePrefix = {arXiv},
       eprint = {1911.12025},
 primaryClass = {astro-ph.HE}
}

@ARTICLE{Loudas2024a,
       author = {{Loudas}, Nick and {Kylafis}, Nikolaos D. and {Tr{\"u}mper}, Joachim},
        title = "{Cyclotron line formation in the radiative shock of an accreting magnetized neutron star}",
      journal = {A\&A},
         year = 2024,
        month = may,
       volume = {685},
          eid = {A95},
        pages = {A95},
          doi = {10.1051/0004-6361/202348109}
}

@Misc{Madsen2020a,
author = {{Madsen}, K.~K. and {Beardmore}, A. and {Page}, K.},
title = {X-calibration of NuSTAR and Swift},
howpublished = {\textsc{url:}~\url{https://iachec.org/wp-content/presentations/2020/Xcal_swift_nustar.pdf}},
month = {11},
year = {2020}
}

@misc{Madsen2021a,
       author = {{Madsen}, K.~K. and {Burwitz}, V. and {Forster}, K. and {Grant}, C.~E. and {Guainazzi}, M. and {Kashyap}, V. and {Marshall}, H.~L. and {Miller}, E.~D. and {Natalucci}, L. and {Plucinsky}, P.~P. and {Terada}, Y.},
        type = "{IACHEC 2020/2021 Pandemic Report}",
         year = 2021,
        month = nov,
          doi = {10.48550/arXiv.2111.01613},
     note = {arXiv:2111.01613}
}

@ARTICLE{Manikantan2023a,
       author = {{Manikantan}, Hemanth and {Paul}, Biswajit and {Rana}, Vikram},
        title = "{An investigation of the '10 keV feature' in the spectra of accretion powered X-ray pulsars with NuSTAR}",
      journal = {MNRAS},
         year = 2023,
        month = nov,
       volume = {526},
       number = {1},
        pages = {1-28},
          doi = {10.1093/mnras/stad2527},
archivePrefix = {arXiv},
       eprint = {2308.15129},
 primaryClass = {astro-ph.HE}
}

@ARTICLE{Manikantan2024a,
       author = {{Manikantan}, Hemanth and {Paul}, Biswajit and {Sharma}, Rahul and {Pradhan}, Pragati and {Rana}, Vikram},
        title = "{Energy dependence of quasi-periodic oscillations in accreting X-ray pulsars}",
      journal = {MNRAS},
         year = 2024,
        month = jun,
       volume = {531},
       number = {1},
        pages = {530-549},
          doi = {10.1093/mnras/stae1170},
archivePrefix = {arXiv},
       eprint = {2404.19323},
 primaryClass = {astro-ph.HE}
}

@ARTICLE{Martin2014,
       author = {{Martin}, Rebecca G. and {Nixon}, Chris and {Armitage}, Philip J. and {Lubow}, Stephen H. and {Price}, Daniel J.},
        title = "{Giant Outbursts in Be/X-Ray Binaries}",
      journal = {ApJ},
         year = 2014,
        month = aug,
       volume = {790},
       number = {2},
          eid = {L34},
        pages = {L34},
          doi = {10.1088/2041-8205/790/2/L34},
archivePrefix = {arXiv},
       eprint = {1407.5676},
 primaryClass = {astro-ph.HE}
}

@ARTICLE{Martin2024,
       author = {{Martin}, Rebecca G. and {Charles}, Philip A.},
        title = "{Disc precession in Be/X-ray binaries drives superorbital variations of outbursts and colour}",
      journal = {MNRAS},
         year = 2024,
        month = feb,
       volume = {528},
       number = {1},
        pages = {L59-L65},
          doi = {10.1093/mnrasl/slad170},
archivePrefix = {arXiv},
       eprint = {2311.06442},
 primaryClass = {astro-ph.HE}
}

@ARTICLE{MeszarosNagel1985a,
       author = {{Meszaros}, P. and {Nagel}, W.},
        title = "{X-ray pulsar models. I. Angle-dependent cyclotron line formation and comptonization.}",
      journal = {ApJ},
         year = 1985,
        month = nov,
       volume = {298},
        pages = {147-160},
          doi = {10.1086/163594}
}

@ARTICLE{MeszarosNagel1985b,
       author = {{Meszaros}, P. and {Nagel}, W.},
        title = "{X-ray pulsar models. II. Comptonized spectra and pulse shapes.}",
      journal = {ApJ},
         year = 1985,
        month = dec,
       volume = {299},
        pages = {138-153},
          doi = {10.1086/163687}
}

@Book{Meszaros1992,
  author    = {{M{\'e}sz{\'a}ros}, P.},
  title     = {{High-energy radiation from magnetized neutron stars.}},
  year      = {1992},
  address = {Chicago},
  publisher = {Univ.\ Chicago Press}
}

@ARTICLE{Miller1989,
       author = {{Miller}, Guy and {Wasserman}, Ira and {Salpeter}, Edwin E.},
        title = "{The Deceleration of Infalling Plasma in Magnetized Neutron Star Atmospheres: Nonisothermal Atmospheres}",
      journal = {ApJ},
         year = 1989,
        month = nov,
       volume = {346},
        pages = {405},
          doi = {10.1086/168020}
}

@ARTICLE{Miyamoto1991a,
       author = {{Miyamoto}, Sigenori and {Kimura}, Kazuhiro and {Kitamoto}, Shunji and {Dotani}, Tadayasu and {Ebisawa}, Ken},
        title = "{X-Ray Variability of GX 339-4 in Its Very High State}",
      journal = {ApJ},
         year = 1991,
        month = dec,
       volume = {383},
        pages = {784},
          doi = {10.1086/170837}
}

@ARTICLE{Mueller2013a,
       author = {{M{\"u}ller}, S. and {Ferrigno}, C. and {K{\"u}hnel}, M. and {Sch{\"o}nherr}, G. and {Becker}, P.~A. and {Wolff}, M.~T. and {Hertel}, D. and {Schwarm}, F. -W. and {Grinberg}, V. and {Obst}, M. and {Caballero}, I. and {Pottschmidt}, K. and {F{\"u}rst}, F. and {Kreykenbohm}, I. and {Rothschild}, R.~E. and {Hemphill}, P. and {N{\'u}{\~n}ez}, S.~M. and {Torrej{\'o}n}, J.~M. and {Klochkov}, D. and {Staubert}, R. and {Wilms}, J.},
        title = "{No anticorrelation between cyclotron line energy and X-ray flux in 4U 0115+634}",
      journal = {A\&A},
         year = 2013,
        month = mar,
       volume = {551},
          eid = {A6},
        pages = {A6},
          doi = {10.1051/0004-6361/201220359},
archivePrefix = {arXiv},
       eprint = {1211.6298},
 primaryClass = {astro-ph.HE}
}

@ARTICLE{Nagel1980,
       author = {{Nagel}, W.},
        title = "{Cyclotron line formation in the accretion column of an X-ray pulsar}",
      journal = {ApJ},
         year = 1980,
        month = mar,
       volume = {236},
        pages = {904-910},
          doi = {10.1086/157817}
}

@ARTICLE{Nagel1981,
       author = {{Nagel}, W.},
        title = "{Radiative transfer in a strongly magnetized plasma. I - Effects of anisotropy. II - Effects of Comptonization}",
      journal = {ApJ},
         year = 1981,
        month = dec,
       volume = {251},
        pages = {278-296},
          doi = {10.1086/159463}
}

@ARTICLE{NagelVentura1983,
       author = {{Nagel}, W. and {Ventura}, J.},
        title = "{Coulomb bremsstrahlung and cyclotron emissivity in hot magnetized plasmas}",
      journal = {A\&A},
         year = 1983,
        month = feb,
       volume = {118},
       number = {1},
        pages = {66-74}
}

@ARTICLE{Nakajima2006a,
       author = {{Nakajima}, M. and {Mihara}, T. and {Makishima}, K. and {Niko}, H.},
        title = "{A Further Study of the Luminosity-dependent Cyclotron Resonance Energies of the Binary X-Ray Pulsar 4U 0115+63 with the Rossi X-Ray Timing Explorer}",
      journal = {ApJ},
         year = 2006,
        month = aug,
       volume = {646},
       number = {2},
        pages = {1125-1138},
          doi = {10.1086/502638},
archivePrefix = {arXiv},
       eprint = {astro-ph/0601491},
 primaryClass = {astro-ph}
}

@ARTICLE{Negueruela2001a,
       author = {{Negueruela}, I. and {Okazaki}, A.~T.},
        title = "{The Be/X-ray transient 4U 0115+63/V635 Cassiopeiae. I. A consistent model}",
      journal = {A\&A},
         year = 2001,
        month = apr,
       volume = {369},
        pages = {108-116},
          doi = {10.1051/0004-6361:20010146},
archivePrefix = {arXiv},
       eprint = {astro-ph/0011407},
 primaryClass = {astro-ph}
}

@ARTICLE{Nelson1995,
       author = {{Nelson}, Robert W. and {Wang}, John C.~L. and {Salpeter}, E.~E. and
         {Wasserman}, Ira},
        title = "{A Potential Cyclotron Line Signature in Low-Luminosity X-Ray Sources}",
      journal = {ApJ},
         year = 1995,
        month = jan,
       volume = {438},
        pages = {L99},
          doi = {10.1086/187725},
archivePrefix = {arXiv},
       eprint = {astro-ph/9410017},
 primaryClass = {astro-ph}
}

@ARTICLE{Okazaki2013,
       author = {{Okazaki}, Atsuo T. and {Hayasaki}, Kimitake and {Moritani}, Yuki},
        title = "{Origin of Two Types of X-Ray Outbursts in Be/X-Ray Binaries. I. Accretion Scenarios}",
      journal = {PASJ},
         year = 2013,
        month = apr,
       volume = {65},
          eid = {41},
        pages = {41},
          doi = {10.1093/pasj/65.2.41},
archivePrefix = {arXiv},
       eprint = {1211.5225},
 primaryClass = {astro-ph.HE}
}

@ARTICLE{Nelson1993,
   author = {{Nelson}, R.~W. and {Salpeter}, E.~E. and {Wasserman}, I.},
    title = "{Nonthermal Cyclotron Emission from Low-Luminosity Accretion onto Magnetic Neutron Stars}",
  journal = {ApJ},
     year = 1993,
    month = dec,
   volume = 418,
    pages = {874},
      doi = {10.1086/173445}
}

@ARTICLE{Pottschmidt2005a,
       author = {{Pottschmidt}, Katja and {Kreykenbohm}, Ingo and {Wilms}, J{\"o}rn and {Coburn}, Wayne and {Rothschild}, Richard E. and {Kretschmar}, Peter and {McBride}, Vanessa and {Suchy}, Slawomir and {Staubert}, R{\"u}diger},
        title = "{RXTE Discovery of Multiple Cyclotron Lines during the 2004 December Outburst of V0332+53}",
      journal = {ApJ},
         year = 2005,
        month = nov,
       volume = {634},
       number = {1},
        pages = {L97-L100},
          doi = {10.1086/498689},
archivePrefix = {arXiv},
       eprint = {astro-ph/0511288},
 primaryClass = {astro-ph}
}

@ARTICLE{Rappaport1978a,
       author = {{Rappaport}, S. and {Clark}, G.~W. and {Cominsky}, L. and {Joss}, P.~C. and {Li}, F.},
        title = "{Orbital elements of 4U 0115+63 and the nature of the hard X-ray transients.}",
      journal = {ApJ},
         year = 1978,
        month = aug,
       volume = {224},
        pages = {L1-L4},
          doi = {10.1086/182745}
}

@ARTICLE{Rouco2017a,
       author = {{Rouco Escorial}, A. and {Bak Nielsen}, A.~S. and {Wijnands}, R. and {Cavecchi}, Y. and {Degenaar}, N. and {Patruno}, A.},
        title = "{The low-luminosity behaviour of the 4U 0115+63 Be/X-ray transient}",
      journal = {MNRAS},
         year = 2017,
        month = dec,
       volume = {472},
       number = {2},
        pages = {1802-1808},
          doi = {10.1093/mnras/stx2111},
archivePrefix = {arXiv},
       eprint = {1704.00284},
 primaryClass = {astro-ph.HE}
}

@ARTICLE{Roy2019a,
       author = {{Roy}, Jayashree and {Agrawal}, P.~C. and {Iyer}, N.~K. and {Bhattacharya}, D. and {Yadav}, J.~S. and {Antia}, H.~M. and {Chauhan}, J.~V. and {Choudhury}, M. and {Dedhia}, D.~K. and {Katoch}, T. and {Madhavani}, P. and {Manchanda}, R.~K. and {Misra}, R. and {Pahari}, M. and {Paul}, B. and {Shah}, P.},
        title = "{LAXPC/AstroSat Study of {\ensuremath{\sim}}1 and {\ensuremath{\sim}}2 mHz Quasi-periodic Oscillations in the Be/X-Ray Binary 4U 0115+63 during Its 2015 Outburst}",
      journal = {ApJ},
         year = 2019,
        month = feb,
       volume = {872},
       number = {1},
          eid = {33},
        pages = {33},
          doi = {10.3847/1538-4357/aafaf1},
archivePrefix = {arXiv},
       eprint = {1901.09382},
 primaryClass = {astro-ph.HE}
}

@article{Roy2024a,
  title = {Luminosity Dependence of the Multiple Cyclotron Lines in {{4U}} 0115+63},
  author = {Roy, Kinjal and Manikantan, Hemanth and Paul, Biswajit},
  year = {2024},
  month = oct,
  volume = {690},
  pages = {A50},
  publisher = {EDP},
  doi = {10.1051/0004-6361/202450395},
  journal = {A\&A}
}

@ARTICLE{Santangelo1999a,
       author = {{Santangelo}, A. and {Segreto}, A. and {Giarrusso}, S. and {Dal Fiume}, D. and {Orlandini}, M. and {Parmar}, A.~N. and {Oosterbroek}, T. and {Bulik}, T. and {Mihara}, T. and {Campana}, S. and {Israel}, G.~L. and {Stella}, L.},
        title = "{A BEPPOSAX Study of the Pulsating Transient X0115+63: The First X-Ray Spectrum with Four Cyclotron Harmonic Features}",
      journal = {ApJ},
         year = 1999,
        month = sep,
       volume = {523},
       number = {1},
        pages = {L85-L88},
          doi = {10.1086/312249}
}

@ARTICLE{Schoenherr2007a,
       author = {{Sch{\"o}nherr}, G. and {Wilms}, J. and {Kretschmar}, P. and {Kreykenbohm}, I. and {Santangelo}, A. and {Rothschild}, R.~E. and {Coburn}, W. and {Staubert}, R.},
        title = "{A model for cyclotron resonance scattering features}",
      journal = {A\&A},
         year = 2007,
        month = sep,
       volume = {472},
       number = {2},
        pages = {353-365},
          doi = {10.1051/0004-6361:20077218},
archivePrefix = {arXiv},
       eprint = {0707.2105},
 primaryClass = {astro-ph}
}

@ARTICLE{Schwarm2017b,
       author = {{Schwarm}, F. -W. and {Sch{\"o}nherr}, G. and {Falkner}, S. and
         {Pottschmidt}, K. and {Wolff}, M.~T. and {Becker}, P.~A. and
         {Sokolova-Lapa}, E. and {Klochkov}, D. and {Ferrigno}, C. and
         {F{\"u}rst}, F. and {Hemphill}, P.~B. and {Marcu-Cheatham}, D.~M. and
         {Dauser}, T. and {Wilms}, J.},
        title = "{Cyclotron resonant scattering feature simulations. I. Thermally averaged cyclotron scattering cross sections, mean free photon-path tables, and electron momentum sampling}",
      journal = {A\&A},
         year = 2017,
        month = jan,
       volume = {597},
          eid = {A3},
        pages = {A3},
          doi = {10.1051/0004-6361/201629352},
archivePrefix = {arXiv},
       eprint = {1609.05030},
 primaryClass = {astro-ph.HE}
}

@phdthesis{SokolovaThesis,
         type = {{PhD thesis}},
        title = {Radiative transfer in the vicinity of accreting neutron stars - polarized emission in strong magnetic fields},
       author = {{Sokolova-Lapa}, Ekaterina },
       school = {Friedrich-Alexander-Universität Erlangen-Nürnberg},
         year = {2023},
          url = {https://www.sternwarte.uni-erlangen.de/docs/theses/2023-11_Sokolova-Lapa.pdf}
}

@ARTICLE{Sokolova2023a,
       author = {{Sokolova-Lapa}, E. and {Stierhof}, J. and {Dauser}, T. and {Wilms}, J.},
        title = "{Vacuum polarization alters the spectra of accreting X-ray pulsars}",
      journal = {A\&A},
         year = 2023,
        month = jun,
       volume = {674},
          eid = {L2},
        pages = {L2},
          doi = {10.1051/0004-6361/202346265},
archivePrefix = {arXiv},
       eprint = {2305.00475},
 primaryClass = {astro-ph.HE}
}

@ARTICLE{Staubert2019a,
       author = {{Staubert}, R. and {Tr{\"u}mper}, J. and {Kendziorra}, E. and {Klochkov}, D. and {Postnov}, K. and {Kretschmar}, P. and {Pottschmidt}, K. and {Haberl}, F. and {Rothschild}, R.~E. and {Santangelo}, A. and {Wilms}, J. and {Kreykenbohm}, I. and {F{\"u}rst}, F.},
        title = "{Cyclotron lines in highly magnetized neutron stars}",
      journal = {A\&A},
         year = 2019,
        month = feb,
       volume = {622},
          eid = {A61},
        pages = {A61},
          doi = {10.1051/0004-6361/201834479},
archivePrefix = {arXiv},
       eprint = {1812.03461},
 primaryClass = {astro-ph.HE}
}

@ARTICLE{Stierhof2025a,
       author = {{Stierhof}, J.~J.~R. and {Sokolova-Lapa}, E. and {Berger}, K. and {Vasilopoulos}, G. and {Thalhammer}, P. and {Zalot}, N. and {Ballhausen}, R. and {El Mellah}, I. and {Malacaria}, C. and {Rothschild}, R.~E. and {Kretschmar}, P. and {Pottschmidt}, K. and {Wilms}, J.},
        title = "{Don't torque like that: Measuring compact object magnetic fields with analytic torque models}",
      journal = {A\&A},
         year = 2025,
        month = jun,
       volume = {698},
          eid = {A308},
        pages = {A308},
          doi = {10.1051/0004-6361/202553809},
archivePrefix = {arXiv},
       eprint = {2504.08700},
 primaryClass = {astro-ph.HE}
}

@misc{Troja2020a,
    organization={Swift Science Center},
    author={{Troja}, Eleonora},
    year={2020},
    title={The Neil Gehrels Swift Observatory, Technical Handbook, Version 17.0},
    howpublished={\url{https://swift.gsfc.nasa.gov/proposals/tech_appd/swiftta_v17.pdf}},
    note={Accessed on 2023-08-17}
}

@ARTICLE{Thalhammer2021a,
       author = {{Thalhammer}, Philipp and {Bissinger}, Matthias and {Ballhausen}, Ralf and {Pottschmidt}, Katja and {Wolff}, Michael T. and {Stierhof}, Jakob and {Sokolova-Lapa}, Ekaterina and {F{\"u}rst}, Felix and {Malacaria}, Christian and {Gottlieb}, Amy and {Marcu-Cheatham}, Diana M. and {Becker}, Peter A. and {Wilms}, J{\"o}rn},
        title = "{Fitting strategies of accretion column models and application to the broadband spectrum of Cen X-3}",
      journal = {A\&A},
         year = 2021,
        month = dec,
       volume = {656},
          eid = {A105},
        pages = {A105},
          doi = {10.1051/0004-6361/202140582},
archivePrefix = {arXiv},
       eprint = {2109.14565},
 primaryClass = {astro-ph.HE}
}

@ARTICLE{Tsygankov2016a,
       author = {{Tsygankov}, S.~S. and {Lutovinov}, A.~A. and {Doroshenko}, V. and {Mushtukov}, A.~A. and {Suleimanov}, V. and {Poutanen}, J.},
        title = "{Propeller effect in two brightest transient X-ray pulsars: 4U 0115+63 and V 0332+53}",
      journal = {A\&A},
         year = 2016,
        month = aug,
       volume = {593},
          eid = {A16},
        pages = {A16},
          doi = {10.1051/0004-6361/201628236},
archivePrefix = {arXiv},
       eprint = {1602.03177},
 primaryClass = {astro-ph.HE}
}

@ARTICLE{Unger1998a,
       author = {{Unger}, S.~J. and {Roche}, P. and {Negueruela}, I. and {Ringwald}, F.~A. and {Lloyd}, C. and {Coe}, M.~J.},
        title = "{Optical spectroscopy of V635 Cassiopeiae/4U 0115+63}",
      journal = {A\&A},
         year = 1998,
        month = aug,
       volume = {336},
        pages = {960-965},
archivePrefix = {arXiv},
       eprint = {astro-ph/9802086},
 primaryClass = {astro-ph}
}

@ARTICLE{Verner1996a,
       author = {{Verner}, D.~A. and {Ferland}, G.~J. and {Korista}, K.~T. and {Yakovlev}, D.~G.},
        title = "{Atomic Data for Astrophysics. II. New Analytic FITS for Photoionization Cross Sections of Atoms and Ions}",
      journal = {ApJ},
         year = 1996,
        month = jul,
       volume = {465},
        pages = {487},
          doi = {10.1086/177435},
archivePrefix = {arXiv},
       eprint = {astro-ph/9601009},
 primaryClass = {astro-ph}
}

@ARTICLE{Vybornov2017a,
       author = {{Vybornov}, V. and {Klochkov}, D. and {Gornostaev}, M. and {Postnov}, K. and {Sokolova-Lapa}, E. and {Staubert}, R. and {Pottschmidt}, K. and {Santangelo}, A.},
        title = "{Luminosity-dependent changes of the cyclotron line energy and spectral hardness in Cepheus X-4}",
      journal = {A\&A},
         year = 2017,
        month = may,
       volume = {601},
          eid = {A126},
        pages = {A126},
          doi = {10.1051/0004-6361/201630275},
archivePrefix = {arXiv},
       eprint = {1702.06361},
 primaryClass = {astro-ph.HE}
}

@ARTICLE{West2017a,
       author = {{West}, Brent F. and {Wolfram}, Kenneth D. and {Becker}, Peter A.},
        title = "{A New Two-fluid Radiation-hydrodynamical Model for X-Ray Pulsar Accretion Columns}",
      journal = {ApJ},
         year = 2017,
        month = feb,
       volume = {835},
       number = {2},
          eid = {129},
        pages = {129},
          doi = {10.3847/1538-4357/835/2/129},
archivePrefix = {arXiv},
       eprint = {1612.02411},
 primaryClass = {astro-ph.HE}
}

@ARTICLE{West2017b,
       author = {{West}, Brent F. and {Wolfram}, Kenneth D. and {Becker}, Peter A.},
        title = "{Dynamical and Radiative Properties of X-Ray Pulsar Accretion Columns: Phase-averaged Spectra}",
      journal = {ApJ},
         year = 2017,
        month = feb,
       volume = {835},
       number = {2},
          eid = {130},
        pages = {130},
          doi = {10.3847/1538-4357/835/2/130},
archivePrefix = {arXiv},
       eprint = {1612.01935},
 primaryClass = {astro-ph.HE}
}

@ARTICLE{West2024a,
       author = {{West}, Brent F. and {Becker}, Peter A. and {Vasilopoulos}, Georgios},
        title = "{Theoretical Analysis of the RX J0209.6‑7427 X-Ray Spectrum during a Giant Outburst}",
      journal = {ApJ},
         year = 2024,
        month = may,
       volume = {966},
       number = {1},
          eid = {L5},
        pages = {L5},
          doi = {10.3847/2041-8213/ad3b92},
archivePrefix = {arXiv},
       eprint = {2404.16202},
 primaryClass = {astro-ph.HE}
}

@ARTICLE{Wijnands2016a,
       author = {{Wijnands}, R. and {Degenaar}, N.},
        title = "{Meta-stable low-level accretion rate states or neutron star crust cooling in the Be/X-ray transients V0332+53 and 4U 0115+63}",
      journal = {MNRAS},
         year = 2016,
        month = nov,
       volume = {463},
       number = {1},
        pages = {L46-L50},
          doi = {10.1093/mnrasl/slw096},
archivePrefix = {arXiv},
       eprint = {1602.02275},
 primaryClass = {astro-ph.HE}
}

@ARTICLE{Wilms2000a,
       author = {{Wilms}, J. and {Allen}, A. and {McCray}, R.},
        title = "{On the Absorption of X-Rays in the Interstellar Medium}",
      journal = {ApJ},
         year = 2000,
        month = oct,
       volume = {542},
       number = {2},
        pages = {914-924},
          doi = {10.1086/317016},
archivePrefix = {arXiv},
       eprint = {astro-ph/0008425},
 primaryClass = {astro-ph}
}

@MISC{XRDB2000,
       AUTHOR = {{Kortright}, Jeffrey B. and {Thompson}, Albert C.},
        TITLE = {X-Ray Data Booklet -- Section 1.2 {X}-ray emission energies},
        YEAR  = {2000},
 HOWPUBLISHED = {\url{https://xdb.lbl.gov/Section1/Sec_1-2.html}},
         NOTE = {Accessed on 2021-06-16}
}

@MISC{XspecManual,
       AUTHOR = {{Arnaud}, Keith and {Gordon}, Craig and {Dorman}, Ben},
        TITLE = {Xspec -- An {X}-Ray Spectral Fitting Package},
        YEAR  = {2022},
        MONTH = {February},
 HOWPUBLISHED = {\url{https://heasarc.gsfc.nasa.gov/xanadu/xspec/manual/XspecManual.html}},
         NOTE = {Accessed on 2022-09-21}
}
\newpage

\begin{appendix}

\onecolumn
\section{Additional figures (N1)}
\label{Appendix_N1}
\begin{figure*}[h!]
 \includegraphics[height=21\baselineskip]{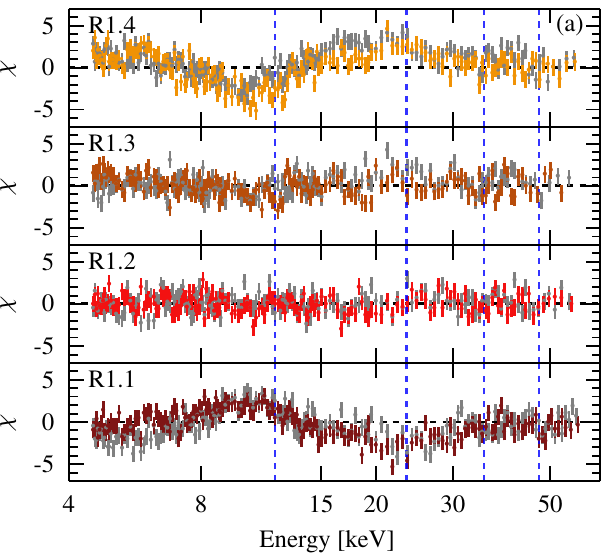}\hfill
 \includegraphics[height=21\baselineskip]{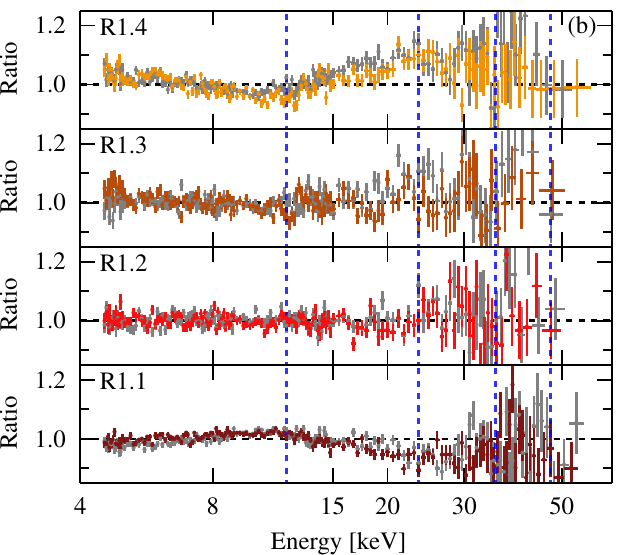}\\
  \includegraphics[height=21\baselineskip]{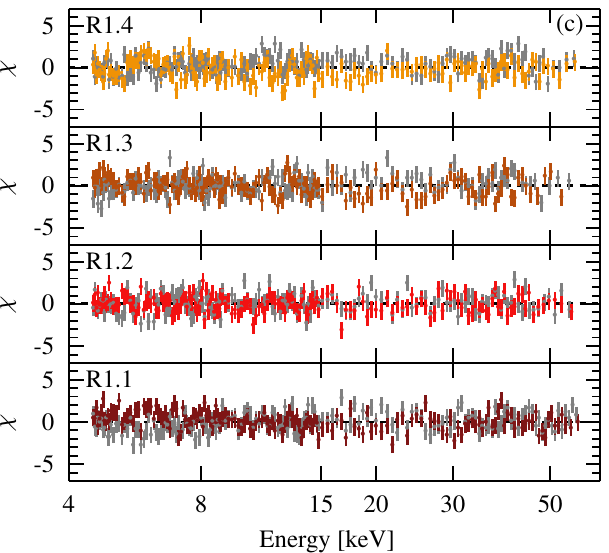}\hfill
 \includegraphics[height=21\baselineskip]{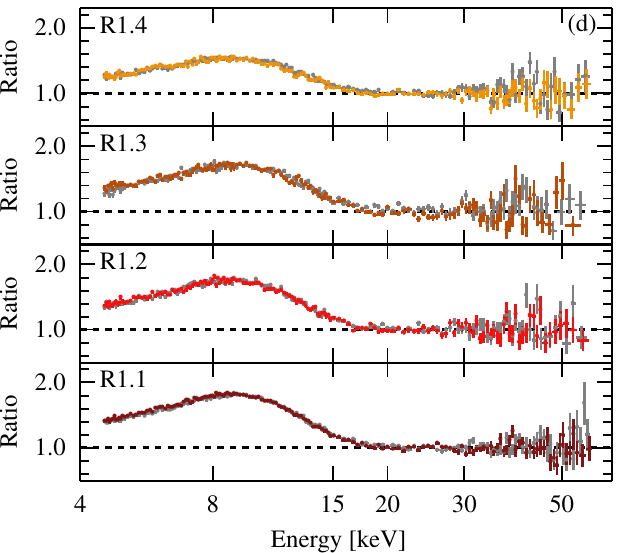}
  \caption{Residuals of the flux-resolved spectra of N1 (FPMA in gray, FPMB in varying color). The blue dashed lines in panels (a) and (b) mark the positions of the CRSFs, using the values listed in Table~\ref{tab:totalPar_NuSTAR_Swift}. 
  The $\chi^2$ residuals (a) and ratio (b), when only evaluating and re-normalizing the flux-averaged best-fit parameters, are shown. Refitting the parameters of the four flux-resolved spectra simultaneously, lead to the best fit as given in panel (c). Panel (d) highlights the contribution of the \keVF, by showing the resulting ratio after removing the component from the best-fit that is shown in the lower left panel.}
 \label{fig:4u2_residuals}
\end{figure*}

In the main body of the paper we showed the results from the flux-resolved analysis of N2, see Fig.~\ref{fig:4u1_residuals}. The residuals of the analysis of observation N1 are shown here in Figure~\ref{fig:4u2_residuals}.

\twocolumn

\renewcommand{\arraystretch}{1.2}

\begin{table*}[h!]

\section{Fit parameters}
\label{Appendix_Parameters}

Table~\ref{tab:totalPar_NuSTAR_Swift} lists all best fit parameters of the spectral fits to both \nustar observations of \fu, both with and without considering the corresponding \swift observations. The best fit parameters for the flux-resolved analysis are listed in Table~\ref{tab:RATE_Par_4ux_compact_20240516}.

  \centering
  \caption{Best parameters of the averaged spectral fit with confidence limits. \label{tab:totalPar_NuSTAR_Swift}}
  \begin{tabular}{lrrrr}
  \hline\hline
  Parameter&N1&O1&N2&O2 \\ 
  \hline
    $c_\mathrm{FPMA}$   & $1.00\tablefootmark{$\dagger$}$ & $1.00\tablefootmark{$\dagger$}$ & $1.00\tablefootmark{$\dagger$}$ & $1.00\tablefootmark{$\dagger$}$ \\ 
    $c_\mathrm{FPMB}$   & $0.9619\pm0.0017$ & $0.9619\pm0.0017$ & $0.9732\pm0.0016$ & $0.9731\pm0.0016$ \\ 
    $c_\mathrm{XRT}$   &  & $0.941\pm0.012$ &  & $0.890\pm0.013$ \\ 
    $N_\mathrm{H}$ [$10^{22}$\,cm$^{-2}$] & $2.3\pm0.8$ & $1.93\pm0.07$ & $1.2\pm0.7$ & $1.99\pm0.07$ \\ 
    $E_\mathrm{CRSF,\,0}$ [keV] & $11.78\pm0.09$ & $11.76\pm0.08$ & $11.76^{+0.11}_{-0.10}$ & $11.71\pm0.10$ \\ 
    $\sigma_\mathrm{CRSF,\,0}$ [keV] & $1.23\tablefootmark{b}$ & $1.25\tablefootmark{b}$ & $1.39\tablefootmark{b}$ & $1.42\tablefootmark{b}$ \\ 
    $D_\mathrm{CRSF,\,0}$ [keV] & $0.173^{+0.028}_{-0.024}$ & $0.182^{+0.025}_{-0.023}$ & $0.134^{+0.034}_{-0.029}$ & $0.155^{+0.036}_{-0.030}$ \\ 
    $E_\mathrm{CRSF,\,1}$ [keV] & $23.6\tablefootmark{a}$ & $23.5\tablefootmark{a}$ & $23.5\tablefootmark{a}$ & $23.4\tablefootmark{a}$ \\ 
    $\sigma_\mathrm{CRSF,\,1}$ [keV] & $2.45\tablefootmark{b}$ & $2.50\tablefootmark{b}$ & $2.78\tablefootmark{b}$ & $2.85\tablefootmark{b}$ \\ 
    $D_\mathrm{CRSF,\,1}$ [keV] & $1.01^{+0.14}_{-0.13}$ & $1.05^{+0.14}_{-0.13}$ & $1.12^{+0.20}_{-0.18}$ & $1.15^{+0.19}_{-0.17}$ \\ 
    $E_\mathrm{CRSF,\,2}$ [keV] & $35.3\tablefootmark{a}$ & $35.3\tablefootmark{a}$ & $35.3\tablefootmark{a}$ & $35.1\tablefootmark{a}$ \\ 
    $\sigma_\mathrm{CRSF,\,2}$ [keV] & $3.68\tablefootmark{b}$ & $3.75\tablefootmark{b}$ & $4.16\tablefootmark{b}$ & $4.27\tablefootmark{b}$ \\ 
    $D_\mathrm{CRSF,\,2}$ [keV] & $1.9\pm0.4$ & $1.9\pm0.4$ & $2.1\pm0.6$ & $2.2^{+0.6}_{-0.5}$ \\ 
    $E_\mathrm{CRSF,\,3}$ [keV] & $47.1\tablefootmark{a}$ & $47.0\tablefootmark{a}$ & $47.0\tablefootmark{a}$ & $46.8\tablefootmark{a}$ \\ 
    $\sigma_\mathrm{CRSF,\,3}$ [keV] & $4.91\tablefootmark{b}$ & $5.01\tablefootmark{b}$ & $5.55\tablefootmark{b}$ & $5.70\tablefootmark{b}$ \\ 
    $D_\mathrm{CRSF,\,3}$ [keV] & $2.1\pm0.9$ & $2.2\pm0.9$ & $< 1.88$ & $1.1^{+1.1}_{-1.0}$ \\ 
    $\Gamma$   & $0.37^{+0.06}_{-0.07}$ & $0.390\pm0.030$ & $0.34\pm0.05$ & $0.40\pm0.04$ \\ 
    $E_\mathrm{fold}$ [keV] & $9.50^{+0.28}_{-0.27}$ & $9.56^{+0.24}_{-0.23}$ & $9.14^{+0.30}_{-0.28}$ & $9.33^{+0.28}_{-0.26}$ \\ 
    $F_\mathrm{PL}$ [keV\,s$^{-1}$\,cm$^{-2}$] & $8.68^{+0.26}_{-0.27}$ & $8.77^{+0.15}_{-0.14}$ & $6.04\pm0.12$ & $6.16\pm0.09$ \\ 
    $A_\mathrm{10\,keV}$ [ph\,s$^{-1}$\,cm$^{-2}$] & $1.00\tablefootmark{$\dagger$}$ & $1.00\tablefootmark{$\dagger$}$ & $1.00\tablefootmark{$\dagger$}$ & $1.00\tablefootmark{$\dagger$}$ \\ 
    $E_\mathrm{10\,keV}$ [keV] & $6.64^{+0.30}_{-0.33}$ & $6.84^{+0.17}_{-0.18}$ & $7.41\pm0.15$ & $7.53\pm0.12$ \\ 
    $\sigma_\mathrm{10\,keV}$ [keV] & $3.83^{+0.18}_{-0.17}$ & $3.75\pm0.11$ & $3.30\pm0.09$ & $3.23\pm0.08$ \\ 
    $F_\mathrm{10\,keV}$ [keV\,s$^{-1}$\,cm$^{-2}$]\tablefootmark{$\triangle$} & $2.69^{+0.29}_{-0.27}$ & $2.5932\pm0.0025$ & $1.69^{+0.11}_{-0.10}$ & $1.6530^{+0.0017}_{-0.0014}$ \\ 
    $F_\mathrm{10\,keV}$/$F_\mathrm{PL}$   & $0.310^{+0.033}_{-0.029}$ & $0.296^{+0.019}_{-0.018}$ & $0.280\pm0.016$ & $0.268\pm0.013$ \\ 
    $A_\mathrm{Fe K\alpha}$ [10$^{-3}$ ph\,s$^{-1}$\,cm$^{-2}$] & $1.4\pm0.4$ & $1.60\pm0.28$ & $1.19\pm0.21$ & $1.13\pm0.19$ \\ 
    $E_\mathrm{Fe K\alpha}$ [keV] & $6.40\tablefootmark{$\dagger$}$ & $6.40\tablefootmark{$\dagger$}$ & $6.40\tablefootmark{$\dagger$}$ & $6.40\tablefootmark{$\dagger$}$ \\ 
    $A_\mathrm{FeXXV\,K\alpha}$ [10$^{-3}$ ph\,s$^{-1}$\,cm$^{-2}$] & $1.02\pm0.30$ & $1.11\pm0.28$ & $1.02^{+0.19}_{-0.20}$ & $0.96\pm0.19$ \\ 
    $E_\mathrm{FeXXV\,K\alpha}$ [keV] & $6.69\tablefootmark{$\dagger$}$ & $6.69\tablefootmark{$\dagger$}$ & $6.69\tablefootmark{$\dagger$}$ & $6.69\tablefootmark{$\dagger$}$ \\ 
    $A_\mathrm{FeXXVI\,K\alpha}$ [10$^{-3}$ ph\,s$^{-1}$\,cm$^{-2}$] & $0.45\pm0.28$ & $0.56\pm0.26$ & $< 0.332$ & $< 0.323$ \\ 
    $E_\mathrm{FeXXVI\,K\alpha}$ [keV] & $6.97\tablefootmark{$\dagger$}$ & $6.97\tablefootmark{$\dagger$}$ & $6.97\tablefootmark{$\dagger$}$ & $6.97\tablefootmark{$\dagger$}$ \\ 
    $c_\mathrm{CRSF,\,\sigma}$   & $0.104^{+0.008}_{-0.007}$ & $0.106\pm0.007$ & $0.118^{+0.011}_{-0.010}$ & $0.122^{+0.011}_{-0.010}$ \\ 
    red. $\chi^2$ / d.o.f. & 1.49 / 326 & 1.50 / 441 & 1.53 / 325  & 1.42 / 439  \\
    \hline
  \end{tabular}
  \tablefoot{The \nustar observations were fitted with (O1 and O2) and without (N1 and N2) using the corresponding \swift observation. The symbols indicate as follows:\\
  \tablefoottext{$\dagger$}{Fixed}\\
  \tablefoottext{a}{Set to the multiple integer value of $E_\mathrm{CRSF,\,0}$}\\
  \tablefoottext{b}{Coupled to the respective CRSF energy via $c_\mathrm{CRSF,\,\sigma}$}\\
  \tablefoottext{$\triangle$}{For fitting, $F_\mathrm{10\,keV}$ is coupled to $F_\mathrm{PL}$ via $F_\mathrm{10\,keV}$/$F_\mathrm{PL}$. The uncertainty of $F_\mathrm{10\,keV}$ is derived by error propagation from the uncertainties of $F_\mathrm{PL}$ and $F_\mathrm{10\,keV}$/$F_\mathrm{PL}$.}}
\end{table*}

\begin{sidewaystable}
\vspace{6cm}
  \caption{Best parameters of the simultaneous spectral fit of the flux-resolved \nustar observations of N1 and N2 with confidence limits.\label{tab:RATE_Par_4ux_compact_20240516}}
  \begin{tabular}{lrrrrrrrr}
  \hline\hline
  Parameter&R11&R12&R13&R14&R21&R22&R23&R24 \\ 
  \hline
    $c_\mathrm{FPMB}\tablefootmark{$\star$}$ & $0.9610\pm0.0017$ & $0.9610\pm0.0017$ & $0.9610\pm0.0017$ & $0.9610\pm0.0017$ & $0.9771\pm0.0016$ & $0.9771\pm0.0016$ & $0.9771\pm0.0016$ & $0.9771\pm0.0016$ \\ 
    $N_\mathrm{H}\tablefootmark{$\star$}$ & $2.7\pm0.8$ & $2.7\pm0.8$ & $2.7\pm0.8$ & $2.7\pm0.8$ & $0.9\pm0.7$ & $0.9\pm0.7$ & $0.9\pm0.7$ & $0.9\pm0.7$ \\ 
    $E_\mathrm{CRSF,\,0}$ & $11.90\pm0.13$ & $11.52^{+0.17}_{-0.18}$ & $11.69^{+0.18}_{-0.19}$ & $11.58^{+0.16}_{-0.17}$ & $12.14^{+0.16}_{-0.15}$ & $11.82^{+0.16}_{-0.17}$ & $11.41^{+0.21}_{-0.22}$ & $11.20\pm0.16$ \\ 
    $\sigma_\mathrm{CRSF,\,0}$ & $1.29\tablefootmark{b}$ & $1.25\tablefootmark{b}$ & $1.27\tablefootmark{b}$ & $1.26\tablefootmark{b}$ & $1.38\tablefootmark{b}$ & $1.34\tablefootmark{b}$ & $1.29\tablefootmark{b}$ & $1.27\tablefootmark{b}$ \\ 
    $D_\mathrm{CRSF,\,0}$ & $0.16\pm0.04$ & $0.17^{+0.06}_{-0.05}$ & $0.26^{+0.07}_{-0.06}$ & $0.23^{+0.06}_{-0.05}$ & $0.10\pm0.04$ & $0.15\pm0.06$ & $0.21^{+0.09}_{-0.08}$ & $0.26\pm0.09$ \\ 
    $E_\mathrm{CRSF,\,1}$ & $23.8\tablefootmark{a}$ & $23.0\tablefootmark{a}$ & $23.4\tablefootmark{a}$ & $23.2\tablefootmark{a}$ & $24.3\tablefootmark{a}$ & $23.6\tablefootmark{a}$ & $22.8\tablefootmark{a}$ & $22.4\tablefootmark{a}$ \\ 
    $\sigma_\mathrm{CRSF,\,1}$ & $2.58\tablefootmark{b}$ & $2.50\tablefootmark{b}$ & $2.54\tablefootmark{b}$ & $2.51\tablefootmark{b}$ & $2.75\tablefootmark{b}$ & $2.68\tablefootmark{b}$ & $2.59\tablefootmark{b}$ & $2.54\tablefootmark{b}$ \\ 
    $D_\mathrm{CRSF,\,1}$ & $1.36^{+0.22}_{-0.20}$ & $1.14^{+0.25}_{-0.23}$ & $0.71^{+0.28}_{-0.26}$ & $0.77^{+0.20}_{-0.19}$ & $1.15^{+0.20}_{-0.17}$ & $1.04^{+0.25}_{-0.22}$ & $0.81^{+0.27}_{-0.26}$ & $1.33^{+0.33}_{-0.30}$ \\ 
    $E_\mathrm{CRSF,\,2}$ & $35.7\tablefootmark{a}$ & $34.6\tablefootmark{a}$ & $35.1\tablefootmark{a}$ & $34.8\tablefootmark{a}$ & $36.4\tablefootmark{a}$ & $35.4\tablefootmark{a}$ & $34.2\tablefootmark{a}$ & $33.6\tablefootmark{a}$ \\ 
    $\sigma_\mathrm{CRSF,\,2}$ & $3.87\tablefootmark{b}$ & $3.75\tablefootmark{b}$ & $3.80\tablefootmark{b}$ & $3.77\tablefootmark{b}$ & $4.13\tablefootmark{b}$ & $4.02\tablefootmark{b}$ & $3.88\tablefootmark{b}$ & $3.81\tablefootmark{b}$ \\ 
    $D_\mathrm{CRSF,\,2}$ & $2.1^{+0.7}_{-0.6}$ & $2.5\pm0.8$ & $2.1\pm1.0$ & $1.6\pm0.7$ & $1.4^{+0.6}_{-0.5}$ & $2.7\pm0.9$ & $2.7\pm1.0$ & $2.8^{+1.2}_{-1.1}$ \\ 
    $E_\mathrm{CRSF,\,3}$ & $47.6\tablefootmark{a}$ & $46.1\tablefootmark{a}$ & $46.8\tablefootmark{a}$ & $46.3\tablefootmark{a}$ & $48.5\tablefootmark{a}$ & $47.3\tablefootmark{a}$ & $45.6\tablefootmark{a}$ & $44.8\tablefootmark{a}$ \\ 
    $\sigma_\mathrm{CRSF,\,3}$ & $5.16\tablefootmark{b}$ & $5.00\tablefootmark{b}$ & $5.07\tablefootmark{b}$ & $5.02\tablefootmark{b}$ & $5.51\tablefootmark{b}$ & $5.36\tablefootmark{b}$ & $5.18\tablefootmark{b}$ & $5.08\tablefootmark{b}$ \\ 
    $D_\mathrm{CRSF,\,3}$ & $2.4^{+1.6}_{-1.5}$ & $4.1^{+2.0}_{-1.9}$ & $2.8^{+2.6}_{-2.5}$ & $2.0^{+1.8}_{-1.7}$ & $< 0.785$ & $3.2^{+2.2}_{-2.0}$ & $< 4.77$ & $< 4.65$ \\ 
    $\Gamma$ & $0.38^{+0.07}_{-0.08}$ & $0.46\pm0.09$ & $0.49\pm0.10$ & $0.44\pm0.08$ & $0.29\pm0.05$ & $0.41\pm0.07$ & $0.43\pm0.08$ & $0.38\pm0.09$ \\ 
    $E_\mathrm{fold}$ & $9.5\pm0.5$ & $10.2^{+0.8}_{-0.7}$ & $10.1^{+1.0}_{-0.8}$ & $9.5\pm0.6$ & $9.04^{+0.25}_{-0.19}$ & $9.4\pm0.6$ & $9.5^{+0.8}_{-0.7}$ & $9.1^{+0.8}_{-0.7}$ \\ 
    $F_\mathrm{PL}$ & $7.01\pm0.24$ & $9.5\pm0.4$ & $10.8\pm0.4$ & $14.0\pm0.4$ & $5.27^{+0.12}_{-0.11}$ & $6.31^{+0.13}_{-0.12}$ & $7.58^{+0.18}_{-0.17}$ & $9.86^{+0.24}_{-0.22}$ \\ 
    $E_\mathrm{10\,keV}\tablefootmark{$\star$}$ & $6.87^{+0.29}_{-0.31}$ & $6.87^{+0.29}_{-0.31}$ & $6.87^{+0.29}_{-0.31}$ & $6.87^{+0.29}_{-0.31}$ & $7.39^{+0.13}_{-0.14}$ & $7.39^{+0.13}_{-0.14}$ & $7.39^{+0.13}_{-0.14}$ & $7.39^{+0.13}_{-0.14}$ \\ 
    $\sigma_\mathrm{10\,keV}$ & $3.75^{+0.19}_{-0.18}$ & $3.66\pm0.19$ & $3.80\pm0.19$ & $3.62\pm0.20$ & $3.45\pm0.10$ & $3.23\pm0.11$ & $3.10^{+0.15}_{-0.16}$ & $2.86^{+0.18}_{-0.19}$ \\ 
    $F_\mathrm{10\,keV}$\tablefootmark{$\triangle$} & $2.31^{+0.28}_{-0.26}$ & $2.8\pm0.4$ & $3.1\pm0.4$ & $3.0\pm0.4$ & $1.77\pm0.12$ & $1.67\pm0.12$ & $1.70\pm0.18$ & $1.57\pm0.20$ \\ 
    $F_\mathrm{10\,keV}$/$F_\mathrm{PL}$ & $0.33\pm0.04$ & $0.29\pm0.04$ & $0.29\pm0.04$ & $0.217^{+0.026}_{-0.024}$ & $0.335\pm0.021$ & $0.265^{+0.019}_{-0.018}$ & $0.224^{+0.023}_{-0.022}$ & $0.159\pm0.020$ \\ 
    $A_\mathrm{Fe\,K\alpha}$ & $1.1\pm0.4$ & $1.5\pm0.8$ & $2.2\pm1.0$ & $2.1\pm0.9$ & $0.86\pm0.22$ & $1.4\pm0.4$ & $2.5\pm0.7$ & $2.7\pm0.9$ \\ 
    $A_\mathrm{FeXXV\,K\alpha}$ & $0.7\pm0.4$ & $1.8^{+0.7}_{-0.8}$ & $< 1.38$ & $1.5\pm0.9$ & $0.88\pm0.21$ & $0.8\pm0.4$ & $1.3\pm0.7$ & $2.3\pm0.9$ \\ 
    $A_\mathrm{FeXXVI\,K\alpha}$ & $< 0.535$ & $< 0.775$ & $1.0\pm0.9$ & $1.3\pm0.9$ & $< 0.0884$ & $0.9\pm0.4$ & $< 0.802$ & $1.1\pm0.9$ \\ 
    $c_\mathrm{CRSF,\,\sigma}\tablefootmark{$\star$}$ & $0.108\pm0.008$ & $0.108\pm0.008$ & $0.108\pm0.008$ & $0.108\pm0.008$ & $0.113\pm0.010$ & $0.113\pm0.010$ & $0.113\pm0.010$ & $0.113\pm0.010$ \\ 
    \hline
  \end{tabular}
  \tablefoot{$N_\mathrm{H}$ is given in units of [$10^{22}$\,cm$^{-2}$],
  $E_\mathrm{*}$, $\sigma_\mathrm{*}$ and $D_\mathrm{*}$ are given in [keV], $A_\mathrm{Fe*}$ have the unit of [10$^{-3}$ ph\,s$^{-1}$\,cm$^{-2}$] and 
  $F_*$ are given in [keV\,s$^{-1}$\,cm$^{-2}$].
  The symbols indicate as follows: 
  \tablefoottext{$\star$}{Tied together for all spectral fits,}
  \tablefoottext{a}{Set to the multiple integer value of $E_\mathrm{CRSF,\,0}$,} \tablefoottext{b}{Coupled to the respective CRSF energy via $c_\mathrm{CRSF,\,\sigma}$,} 
  \tablefoottext{$^\triangle$}{Coupled to $F_\mathrm{PL}$ via $F_\mathrm{10\,keV}$/$F_\mathrm{PL}$. The uncertainty of $F_\mathrm{10\,keV}$ is derived by error propagation from the uncertainties of $F_\mathrm{PL}$ and $F_\mathrm{10\,keV}$/$F_\mathrm{PL}$.}
  }
\end{sidewaystable}

\renewcommand{\arraystretch}{1}

\end{appendix}

\end{document}